\documentclass[8pt]{article}
\usepackage{amsfonts}
\usepackage{latexsym,amsmath}
\usepackage{amssymb,array}
\usepackage{graphics}
\makeatletter \oddsidemargin 0in \evensidemargin 0in \textwidth
16cm \RequirePackage[dvips]{graphicx} \textheight 20cm
\begin{document}
\title{A Generalized One Parameter Polynomial Exponential Generator Family of Distributions}
\author{SUDHANSU S. MAITI$^1$\footnote{Corresponding author. e-mail:dssm1@rediffmail.com} and SUKANTA PRAMANIK$^2$\\
$^1$Department of Statistics, Visva-Bharati University, Santiniketan-731 235, India\\
$^2$Department of Statistics, Siliguri College, North Bengal University, Siliguri-734 001, India}
\date{}
\maketitle

\textbf{Abstract:} A new class of distributions, called Generalized One Parameter Polynomial Exponential-G family of distributions is proposed for modelling lifetime data. An account of the structural and reliability properties of the new class is presented. Maximum likelihood estimation of parameters of the class of distributions has been described. Simulation study results have been reported. Two data sets have been analyzed to illustrate its applicability.\\

\section{Introduction}
Fitting of a probability distribution to real life data and synthesis of information from it becomes more challenging task to the researchers. Classical probability distributions like binomial, Poisson, normal are not sufficient to elicit information properly from data. Data generated from day to day work environment are more complex in nature now-a-days. Statistical distributions are important for parametric inferences and applications to fit real world phenomena.\\
Some methods are developed in the early days for generating univariate continuous distributions like Pearsonian system of distributions by Prof. Karl Pearson[27], Johnson system by Johnson[14], and methods based on quantile functions developed by Tukey[32]. McDonald[24], Azzalini[3], Marshall and Olkin[23] also proposed some general methods for generating a new family of distributions. In this century, Eugene, Lee, and Famoye[12] proposed the beta-generated family of distributions, Jones[15], and Cordeiro and deCastro[9] extended the beta-generated family of distributions by using Kumaraswamy distribution in place of beta distribution.\\
Alzaatreh, Lee, and Famoye[2] proposed a generalized family of distributions, called T-X (also called Transformed-Transformer) family, whose cumulative distribution function (cdf) is given by
\begin{eqnarray}\label{eq1}
	F(x; \theta)= \int_{a}^{W[G(x)]}r(t)dt,
\end{eqnarray}
where, the random variable $T\in [a,b]$, for $-\infty<a,b<\infty$ and $W[G(x)]$ be a function of the cdf $G(x)$ so that $W[G(x)]$ satisfies the following conditions:
\begin{enumerate}
	\item[(i)] $W[G(x)]\in [a,b]$,
	\item[(ii)] $W[G(x)]$ is differentiable and monotonically non-decreasing,
	\item[(iii)] $W[G(x)]\rightarrow a$ as $x\rightarrow -\infty$ and $W[G(x)]\rightarrow b$ as $x\rightarrow \infty$.
\end{enumerate}
In recent years, the Lindley distribution gets popularity over the exponential distribution for its flexibility on some properties like the mode, moments, skewness and kurtosis measures, cumulants, failure rate and mean residual life, mean deviation, entropies, etc. The Lindley distribution is the mixture of an Exponential distribution and a Gamma distribution with shape parameter $2$.\\
Let X is a random variable taking values $(0,\infty)$. So the distribution of X may be absolutely
continuous or discrete. The probability density function (pdf) of Lindley distribution [see,
Lindley (1958)]is given by
\begin{eqnarray*}
	f(x;\lambda)= \frac{\lambda^2}{1+\lambda}(1+x)e^{-\lambda x},~~~~~~~~\lambda,x > 0.
\end{eqnarray*}	
It has been generalized by host of authors. To mention a few, Zakerzadeh and Dolati(2010), Bakouch et al. (2012), Shanker et al. (2013), Elbatal et al. (2013), Ghitany et al. (2013), Singh et al. (2014), Abouamoh et al. (2015) among others. Bouchahed and Zeghdoudi (2018) has proposed a new and unified approach in generalizing the Lindley's distribution. They investigated some structural properties like moments, skewness, kurtosis, median, mean deviations, Lorenz curve, entropies and limiting distribution of extreme order statistics; reliability properties like reliability function, hazard rate, stress-strength reliability, stochastic ordering; and estimation methods like method of moment and maximum likelihood. Bhattacharya et al.(2020)has derived uniform minimum variance unbiased estimators (UMVUEs)of reliability functions (both mission time and stress-strength) and their associated variances.  The probability density function of the random variable X in a one parameter polynomial exponential (abbreviation, OPPE) family can be written as
\begin{eqnarray}\label{eq2}
f_{X}(x;\lambda)= \frac{\sum_{k=0}^{s}a_{k}x^{k}e^{-\lambda x}}{\sum_{k=0}^{s}a_{k}\frac{\Gamma(k+1)}{\lambda^{k+1}}},~~~~~~~~\lambda,x > 0.
\end{eqnarray}	
The distribution can also be written as
\begin{eqnarray*} 
f_{X}(x;\lambda)= h(\lambda)\sum_{k=0}^{s}a_{k}x^{k} e^{-\lambda x}=h(\lambda)\sum_{k=0}^{s}a_{k}	\frac{\Gamma(k+1)}{\lambda^{k+1}}f_{GA}(x;k+1,\lambda)
\end{eqnarray*}
where, $h(\lambda)=\frac{1}{\sum_{k=0}^{s}a_{k}\frac{\Gamma(k+1)}{\lambda^{k+1}}}$, and $f_{GA}(x;k+1,\lambda)$ is the pdf of a gamma distribution with shape parameter $(k+1)$ and scale
parameter $\lambda$, and $a_{k}$'s are non-negative constants. The distribution is a finite mixture of $(s+1)$ gamma distributions.\\
Special cases are\\
(a) $s=0$, $a_0=1$ gives the Exponential distribution,\\
(b) $s=1$, $a_0=1$, $a_1=1$ gives the Lindley distribution,\\
(c) $s=2$, $a_0=1$, $a_1=0$, $a_2=1$ gives the Akash distribution [c.f. Shankar(2015a)],\\
(d) $s=2$, $a_0=1$, $a_1=2$, $a_2=1$ gives the Aradhana distribution [c.f. Shankar(2016a)],\\
(e) $s=2$, $a_0=1$, $a_1=1$, $a_2=1$ gives the Sujatha distribution [c.f. Shankar(2016b)],\\
(f) $s=2$, $a_0=0$, $a_1=1$, $a_2=1$ gives the length-biased Lindley distribution [c.f. Ayesha(2017)],\\
(g) $s=3$, $a_0=1$, $a_1=1$, $a_2=1$, $a_3=1$ gives the Amarendra distribution [c.f. Shankar(2016c)],\\
(h) $s=4$, $a_0=1$, $a_1=1$, $a_2=1$, $a_3=1$, $a_4=1$ gives the Devya distribution [c.f. Shankar(2016d)],\\
(i) $s=5$, $a_0=1$, $a_1=1$, $a_2=1$, $a_3=1$, $a_4=1$, $a_5=1$ gives the Shambhu distribution [c.f. Shankar(2016e)].\\
In this paper, we propose a new wider class of continuous distributions called the Odds One Parameter Polynomial Exponential - G family by taking $W[G(x)]=\frac{G(x;\xi)}{1-G(x;\xi)}$, the odds function of cdf and $r(t)=h(\lambda)\sum_{k=0}^{s}a_{k}t^{k}e^{-\lambda t}, t > 0, \lambda > 0$, the generator. Here $G(x;\xi)$ is a baseline cdf, which depends on a parameter vector $\xi$ and $\bar{G}(x;\xi)= 1- G(x;\xi)$ is the baseline survival function. Throughout this paper we use the following notations. We write upper incomplete gamma function and lower incomplete gamma function as $\Gamma (p,x)=\int_x^{\infty}w^{p-1}e^{-w}dw$ and $\gamma (p,x)=\int_0^{x}w^{p-1}e^{-w}dw$, for $x\geq 0, p>0$ respectively. The j-th derivative with respect to $p$ is denoted by $\Gamma^{(j)} (p,x)=\int_x^{\infty}(\ln w)^jw^{p-1}e^{-w}dw$ and $\gamma^{(j)} (p,x)=\int_0^{x}(\ln w)^jw^{p-1}e^{-w}dw$, for $x\geq 0, p>0$ respectively. Assuming the Exponential distribution as the generator, Maiti and Pramanik[19-21] have developed Odds Generalized Exponential-Exponential, Exponential-Uniform and Exponential-Pareto distributions and their properties studied and applications illustrated. Assuming the xgamma distribution as the generator, Maiti and Pramanik[22] have developed a Odds xgamma - G family of distributions.\\
The distribution function of Odds OPPE - G family of distributions is given by
\begin{eqnarray}\label{eq3}
	F(x;\lambda,\xi)&=& \int_{0}^{\frac{G(x;\xi)}{1-G(x;\xi)}}h(\lambda)\sum_{k=0}^{s}a_{k}t^{k} e^{-\lambda t}dt
	\nonumber\\&=&1-h(\lambda)\sum_{k=0}^{s}a_{k}\frac{\Gamma\left(k+1, \lambda \frac{G(x;\xi)}{\bar{G}(x;\xi)}\right)}{\lambda^{k+1}} 
\end{eqnarray}
where, $h(\lambda)=\frac{1}{\sum_{k=0}^{s}a_{k}\frac{\Gamma(k+1)}{\lambda^{k+1}}}$\\
The probability density function (pdf) of Odds OPPE - G family of distribution is
\begin{eqnarray}\label{eq4}
	f(x;\lambda,\xi)= h(\lambda)\sum_{k=0}^{s}a_{k}\frac{g(x;\xi)}{[\bar{G}(x;\xi)]^2}\left[\frac{G(x;\xi)}{\bar{G}(x;\xi)}\right]^{k}e^{-\lambda \left[\frac{G(x;\xi)}{\bar{G}(x;\xi)}\right]}.
\end{eqnarray}

The survival function of Odds OPPE - G family of distributions is 
\begin{eqnarray}\label{eq5}
	S(x;\lambda,\xi)&=&h(\lambda)\sum_{k=0}^{s}a_{k}\frac{\Gamma\left(k+1, \lambda\frac{G(x;\xi)}{\bar{G}(x;\xi)}\right)}{\lambda^{k+1}}.
\end{eqnarray}

The hazard rate function of Odds OPPE - G family of distribution is 
\begin{eqnarray}\label{eq6}
	h(t;\lambda,\xi)&=&\frac{f(t;\lambda,\xi)}{S(t;\lambda,\xi)}\nonumber\\&=&\frac{\sum_{k=0}^{s}a_{k}\frac{g(t;\xi)}{[\bar{G}(t;\xi)]^2}\left[\frac{G(t;\xi)}{\bar{G}(t;\xi)}\right]^{k}e^{-\lambda \left[\frac{G(t;\xi)}{\bar{G}(t;\xi)}\right]}}{\sum_{k=0}^{s}a_{k}\frac{\Gamma\left(k+1, \lambda\frac{G(t;\xi)}{\bar{G}(t;\xi)}\right)}{\lambda^{k+1}}}.
\end{eqnarray}

Odds function for different distributions and parameter vector $\xi$ have been presented in Table 1. The rest of the article has been organised as follows. Section 2 discusses some particular models assuming transformer distribution as Uniform, Exponential and Burr XII. Section 3 discusses some mathematical properties like Mixture Representation, Shape, Quantile function, Entropy, Order Statistics, Stress-Strength reliability, Incomplete moments, Mean deviations, Lorenz and Bonferroni curves, Moments of residual and reversed residual life. Maximum likelihood method of estimating parameters has been discussed in section 4. Simulation study method has been described and simulation results have been represented in section 5. Application of this model for two data sets have been discussed and reported in section 6. Concluding remarks have been made in section 7.

\begin{table}
\caption{Distributions and corresponding $G(x;\xi)/\bar{G}(x;\xi)$ functions}
\label{tab:1} 
\begin{center}
\begin{tabular}{lll}
\hline\noalign{\smallskip}
Distribution & $G(x;\xi)/\bar{G}(x;\xi)$ & $\xi$  \\
\noalign{\smallskip}\hline\noalign{\smallskip}
Uniform($0<x<\theta$) & $x/(\theta-x)$ & $\theta$ \\
Exponential($x>0$) & $e^{\lambda x}-1$ & $\lambda$ \\
Weibull($x>0$) & $e^{\lambda x^{\gamma}}-1$ & $(\lambda, \gamma)$ \\
Frechet($x>0$) & $(e^{\lambda x^{\gamma}}-1)^{-1}$ & $(\lambda, \gamma)$ \\
Half-logistic($x>0$) & $(e^{x}-1)/2$ & $\phi$ \\
Power function($0<x<1/\theta$) & $[(\theta x)^{-k}-1]^{-1}$ & $(\theta,k)$ \\
Pareto($x\geq \theta$) & $(x/\theta)^k-1$ & $(\theta,k)$ \\
Burr XII($>0$) & $[1+(x/s)^c]^{k}-1$ & $(s,k,c)$\\
Log-logistic($x>0$) & $[1+(x/s)^c]-1$ & $(s,c)$\\
Lomax($x>0$) & $[1+(x/s)]^{k}-1$ & $(s,k)$\\
Gumbel($-\infty<x<\infty$) & $[exp[exp(-(x-\mu)/\sigma)]-1]^{-1}$ & $(\mu,\sigma)$\\
Kumaraswamy($0<x<1$) & $(1-x^a)^{-b}-1$ & $(a,b)$\\
Normal($-\infty<x<\infty$) & $\Phi((x-\mu)/\sigma)/(1-\Phi((x-\mu)/\sigma))$ & $(\mu,\sigma)$\\    
\noalign{\smallskip}\hline
\end{tabular}
\end{center}
\end{table}

\section{Some Special Models for Odds OPPE - G Family}
In this section, some new special distributions, namely, Odds OPPE-Uniform, Odds OPPE-Exponential, Odds OPPE-Pareto, and Odds OPPE-Burr XII are introduced.
\subsection{Odds OPPE - Uniform Distribution}
Consider the baseline distribution as uniform on the interval $(0,\theta),~\theta>0~$with the pdf and cdf, respectively
\begin{eqnarray*}
g(x;\theta)=\frac{1}{\theta}~;0<x<\theta<\infty, ~G(x,\theta)=\frac{x}{\theta}.
\end{eqnarray*}
The cdf of Odds OPPE-Uniform distribution is obtained by substituting the cdf of uniform in $(\ref{eq3})$ as follows
\begin{eqnarray*}
F(x;\lambda,\theta)&=&1-h(\lambda)\sum_{k=0}^{s}a_{k}\frac{\Gamma\left(k+1, \frac{\lambda x}{\theta-x}\right)}{\lambda^{k+1}}.
\end{eqnarray*}
where, $h(\lambda)=\frac{1}{\sum_{k=0}^{s}a_{k}\frac{\Gamma(k+1)}{\lambda^{k+1}}}$\\
The corresponding pdf is given by
\begin{eqnarray*}
f(x;\lambda,\theta)&=&h(\lambda)\sum_{k=0}^{s}a_{k}\frac{\theta}{(\theta-x)^2}\left(\frac{x}{\theta-x}\right)^{k}e^{-\frac{\lambda x}{\theta-x}} ~~;0<x<\theta<\infty, \lambda>0.
\end{eqnarray*}
The survival and hazard rate functions are given respectively as follows:
\begin{eqnarray*}
S(x;\lambda,\theta)&=&h(\lambda)\sum_{k=0}^{s}a_{k}\frac{\Gamma\left(k+1, \frac{\lambda x}{\theta-x}\right)}{\lambda^{k+1}},
\end{eqnarray*}
\begin{eqnarray*}
r(t;\lambda,\theta)&=&\frac{\sum_{k=0}^{s}a_{k}\frac{\theta}{(\theta-t)^2}\left(\frac{t}{\theta-t}\right)^{k}e^{-\frac{\lambda t}{\theta-t}}}{\sum_{k=0}^{s}a_{k}\frac{\Gamma\left(k+1, \frac{\lambda t}{\theta-t}\right)}{\lambda^{k+1}}}.
\end{eqnarray*}
\textbf{Odds Lindley - Uniform Distribution:-}\\
In equation number $(\ref{eq2})$, when $s=1$, $a_0=1$, $a_1=1$, the One Parameter Polynomial Exponential gives the Lindley distribution. So when $s=1$, $a_0=1$, $a_1=1$, the Odds OPPE - Uniform Distribution reduces to Odds Lindley - Uniform Distribution with pdf 
\begin{eqnarray*}
f(x;\lambda,\theta)&=&\frac{\lambda^2}{1+\lambda}.\frac{\theta^2}{(\theta-x)^3} e^{-\frac{\lambda x}{\theta-x}} ~~;0<x<\theta<\infty, \lambda>0.
\end{eqnarray*}

\begin{figure*}
  \includegraphics[width=1\textwidth]{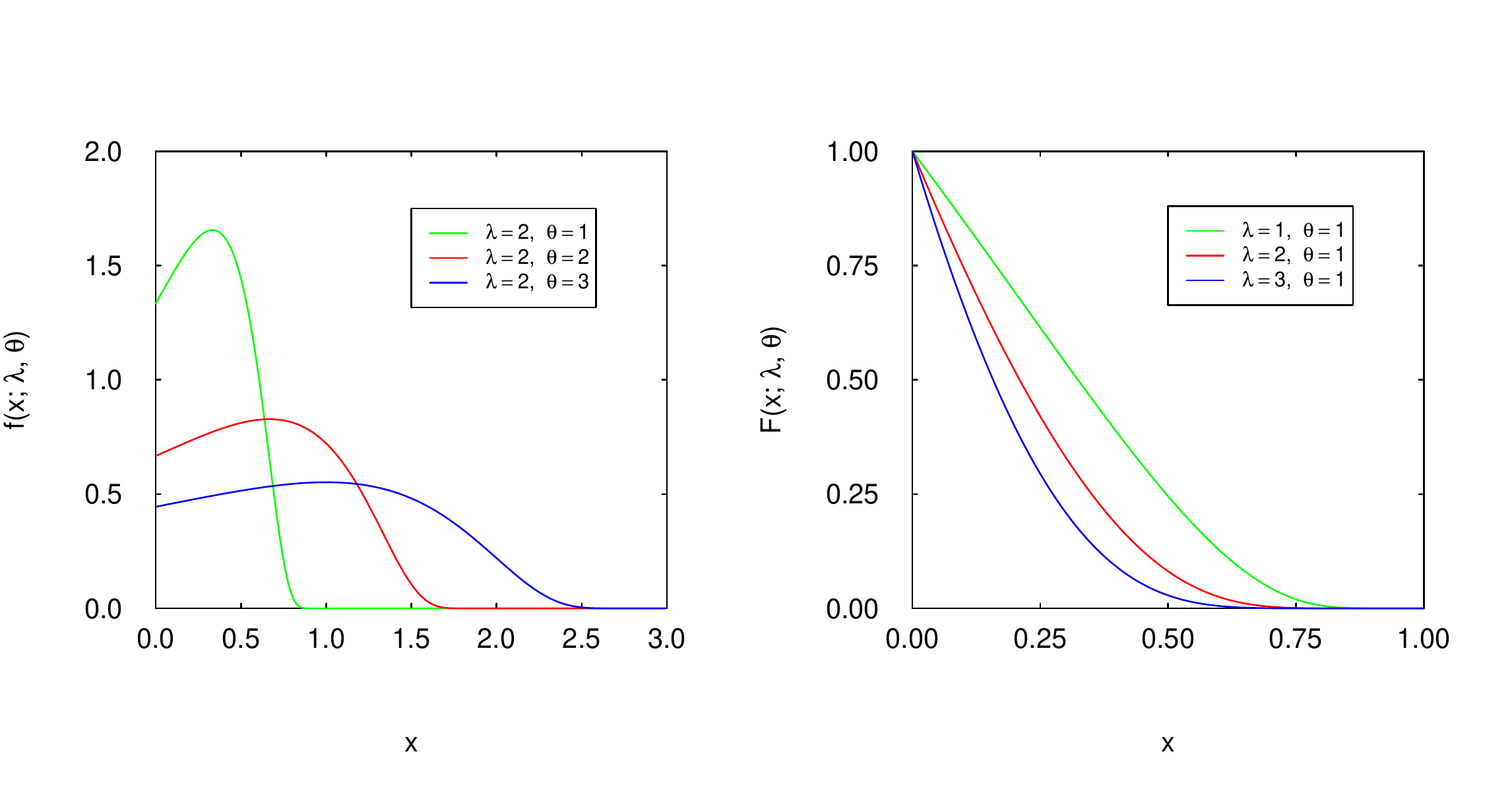}
\caption{The pdf and survival function of Odds Lindley - Uniform distribution}
\label{fig1}
\end{figure*}

\subsection{Odds OPPE - Exponential Distribution}
Considering the baseline distribution is Exponential with parameter $\theta>0~$. The pdf and cdf are
\begin{eqnarray*}
g(x;\theta)=\theta e^{-\theta x}~;0<x,\theta<\infty, ~G(x,\theta)=1-e^{-\theta x}.
\end{eqnarray*}
The cdf of Odds OPPE-Exponential distribution is obtained by substituting the cdf of Exponential in $(\ref{eq3})$ as follows
\begin{eqnarray*}
F(x;\lambda,\theta)&=&1-h(\lambda)\sum_{k=0}^{s}a_{k}\frac{\Gamma\left(k+1, \lambda(e^{\theta x}-1)\right)}{\lambda^{k+1}}.
\end{eqnarray*}
where, $h(\lambda)=\frac{1}{\sum_{k=0}^{s}a_{k}\frac{\Gamma(k+1)}{\lambda^{k+1}}}$\\
The corresponding pdf is given by
\begin{eqnarray*}
f(x;\lambda,\theta)&=&h(\lambda)\sum_{k=0}^{s}a_{k}\theta e^{\theta x}(e^{\theta x}-1)^{k}e^{-\lambda (e^{\theta x}-1)} ~~;0<x,\theta<\infty, \lambda>0.
\end{eqnarray*}
The survival and hazard rate functions are as follows:
\begin{eqnarray*}
S(x;\lambda,\theta)&=&h(\lambda)\sum_{k=0}^{s}a_{k}\frac{\Gamma\left(k+1, \lambda(e^{\theta x}-1)\right)}{\lambda^{k+1}},
\end{eqnarray*}
\begin{eqnarray*}
r(t;\lambda,\theta)&=&\frac{\sum_{k=0}^{s}a_{k}\theta e^{\theta t}(e^{\theta t}-1)^{k}e^{-\lambda (e^{\theta t}-1)}}{\sum_{k=0}^{s}a_{k}\frac{\Gamma\left(k+1, \lambda(e^{\theta t}-1)\right)}{\lambda^{k+1}}}.
\end{eqnarray*}
\textbf{Odds Lindley - Exponential Distribution:-}\\
In equation number $(\ref{eq2})$, when $s=1$, $a_0=1$, $a_1=1$, the One Parameter Polynomial Exponential gives the Lindley distribution. So when $s=1$, $a_0=1$, $a_1=1$, the Odds OPPE - Exponential Distribution reduces to Odds Lindley - Exponential Distribution with pdf 
\begin{eqnarray*}
f(x;\lambda,\theta)&=&\frac{\lambda^2}{(1+ \lambda)}\theta e^{2\theta x}e^{-\lambda(e^{\theta x}-1)} ~~;0<x,\theta<\infty, \lambda>0.
\end{eqnarray*}

\begin{figure*}
  \includegraphics[width=1\textwidth]{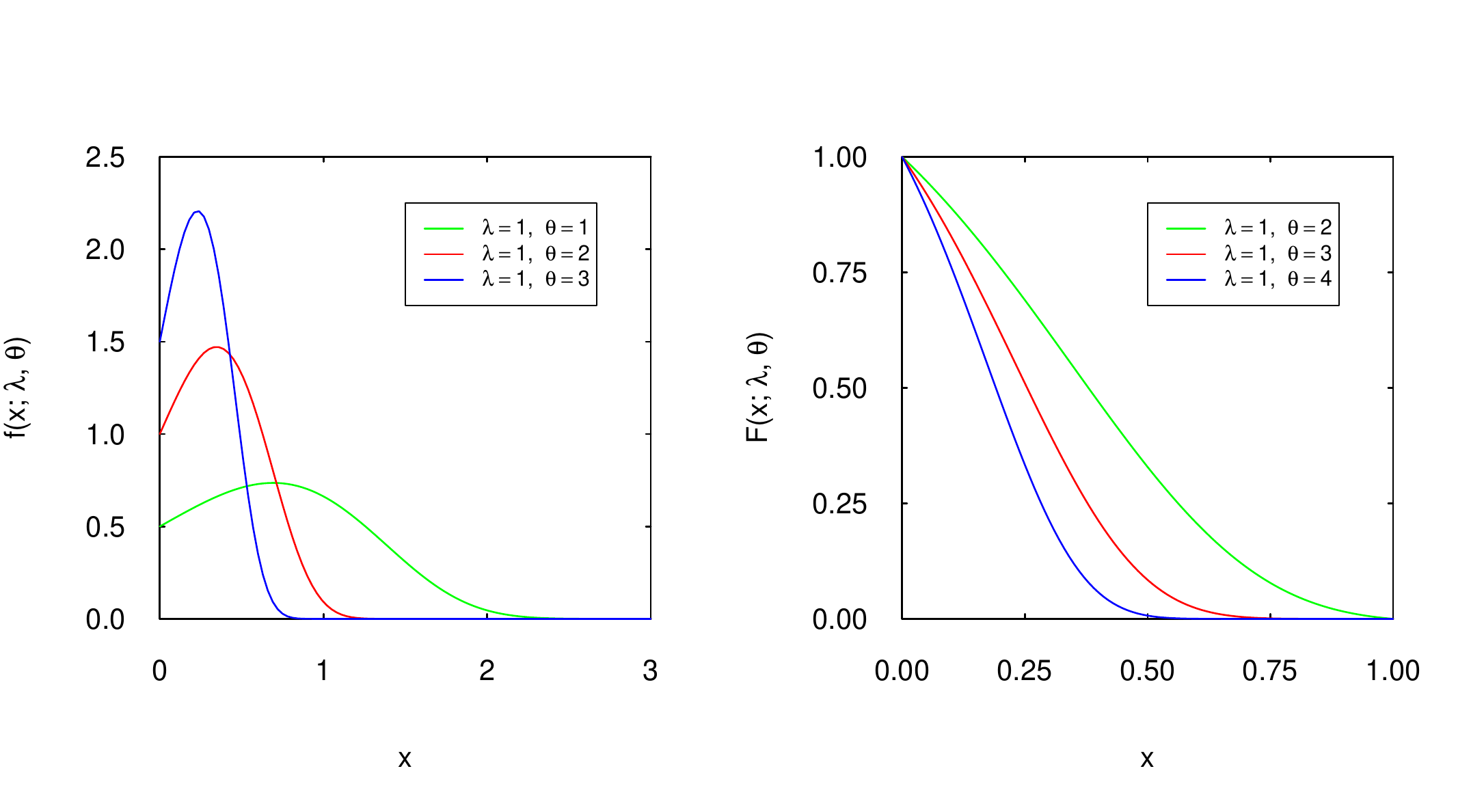}
\caption{The pdf and survival function of Odds Lindley - Exponential distribution}
\label{fig2}
\end{figure*}

\subsection{Odds OPPE - Pareto Distribution}
Considering the baseline distribution is Pareto with parameters $a$, and $\theta>0~$. The pdf and cdf are
\begin{eqnarray*}
g(x;\theta, a)=\frac{\theta a^\theta}{x^{\theta+1}}~;a<x<\infty,\theta>0, ~G(x,\theta, a)=1-\left( \frac{a}{x}\right) ^\theta.
\end{eqnarray*}
The cdf of Odds OPPE-Pareto distribution is obtained by substituting the cdf of Pareto in $(\ref{eq3})$ as follows
\begin{eqnarray*}
F(x;\lambda,\theta,a)&=&1-h(\lambda)\sum_{k=0}^{s}a_{k}\frac{\Gamma\left(k+1, \lambda[\left( \frac{x}{a}\right)^{\theta} -1] \right)}{\lambda^{k+1}}.
\end{eqnarray*}
where, $h(\lambda)=\frac{1}{\sum_{k=0}^{s}a_{k}\frac{\Gamma(k+1)}{\lambda^{k+1}}}$\\
The corresponding pdf is given by
\begin{eqnarray*}
f(x;\lambda,\theta,a)&=&h(\lambda)\sum_{k=0}^{s}a_{k} \frac{\theta x^{\theta-1}}{a^{\theta}}\left\lbrace \left( \frac{x}{a}\right)^{\theta} -1\right\rbrace ^{k}e^{-\lambda[\left( \frac{x}{a}\right)^{\theta} -1]} ~~;a<x<\infty,\theta>0,\lambda>0.
\end{eqnarray*}
The survival and hazard rate functions are as follows:
\begin{eqnarray*}
S(x;\lambda,\theta,a)&=&h(\lambda)\sum_{k=0}^{s}a_{k}\frac{\Gamma\left(k+1, \lambda[\left( \frac{x}{a}\right)^{\theta} -1] \right)}{\lambda^{k+1}},
\end{eqnarray*}
\begin{eqnarray*}
r(t;\lambda,\theta,a)&=&\frac{\sum_{k=0}^{s}a_{k} \frac{\theta t^{\theta-1}}{a^{\theta}}\left\lbrace \left( \frac{t}{a}\right)^{\theta} -1\right\rbrace ^{k}e^{-\lambda[\left( \frac{t}{a}\right)^{\theta} -1]}}{\sum_{k=0}^{s}a_{k}\frac{\Gamma\left(k+1, \lambda[\left( \frac{t}{a}\right)^{\theta} -1] \right)}{\lambda^{k+1}}}.
\end{eqnarray*}
\textbf{Odds Lindley - Pareto Distribution:-}\\
In equation number $(\ref{eq2})$, when $s=1$, $a_0=1$, $a_1=1$, the One Parameter Polynomial Exponential gives the Lindley distribution. So when $s=1$, $a_0=1$, $a_1=1$, the Odds OPPE - Pareto Distribution reduces to Odds Lindley - Pareto Distribution with pdf 
\begin{eqnarray*}
f(x;\lambda,\theta,a)&=&\frac{\lambda^2}{(1+ \lambda)}\frac{\theta x^{2\theta-1}}{a^{2\theta}} e^{-\lambda((\frac{x}{a})^\theta-1)} ~~;a<x<\infty,\theta>0,\lambda>0.
\end{eqnarray*}

\begin{figure*}
  \includegraphics[width=1\textwidth]{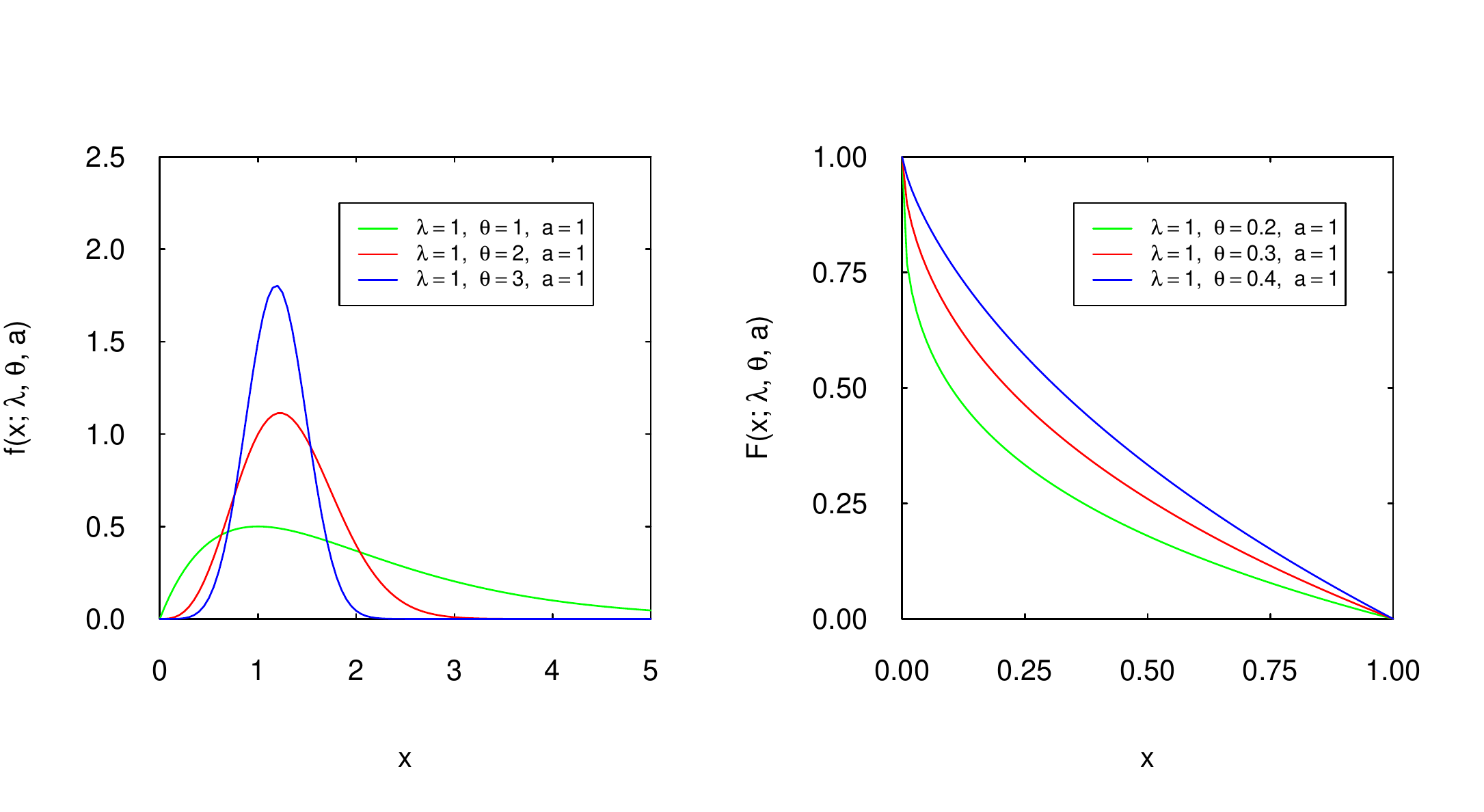}
\caption{The pdf and survival function of Odds Lindley - Pareto distribution}
\label{fig3}
\end{figure*}

\subsection{Odds OPPE - Burr XII Distribution}
Considering the baseline distribution is Burr[6] with the following pdf and cdf
\begin{eqnarray*}
g(x;\alpha,\theta)=\alpha\theta x^{(\alpha-1)}\left(1+x^\alpha\right)^{-(\theta+1)}~~;x\geq0,\alpha,\theta>0, 
\end{eqnarray*}
\begin{eqnarray*}
G(x;\alpha,\theta)=1- \left(1+x^\alpha\right)^{-\theta}~~;x\geq0,\alpha,\theta>0.
\end{eqnarray*}
The cdf of Odds xgamma-Burr XII distribution is obtained by substituting the cdf of Burr XII in $(\ref{eq3})$ as follows
\begin{eqnarray*}
F(x;\lambda,\alpha,\theta)&=&1-h(\lambda)\sum_{k=0}^{s}a_{k}\frac{\Gamma\left(k+1, \lambda[(1+x^{\alpha})^{\theta}-1]\right)}{\lambda^{k+1}}.
\end{eqnarray*}
where, $h(\lambda)=\frac{1}{\sum_{k=0}^{s}a_{k}\frac{\Gamma(k+1)}{\lambda^{k+1}}}$\\
The corresponding pdf is given by
\begin{eqnarray*}
f(x;\lambda,\alpha,\theta)&=&h(\lambda)\sum_{k=0}^{s}a_{k}\alpha \theta x^{\alpha-1}(1+x^{\alpha})^{\theta-1}\left[ (1+x^{\alpha})^{\theta}-1\right]^{k}e^{-\lambda[(1+x^{\alpha})^{\theta}-1]}; \nonumber\\&& 0<x,\theta,\alpha<\infty, \lambda>0.
\end{eqnarray*}
The survival and hazard rate functions are given as follows:
\begin{eqnarray*}
S(x;\lambda,\alpha,\theta)&=&h(\lambda)\sum_{k=0}^{s}a_{k}\frac{\Gamma\left(k+1, \lambda[(1+x^{\alpha})^{\theta}-1]\right)}{\lambda^{k+1}},
\end{eqnarray*}
\begin{eqnarray*}
r(t;\lambda,\alpha,\theta)&=&\frac{\sum_{k=0}^{s}a_{k}\alpha \theta t^{\alpha-1}(1+t^{\alpha})^{\theta-1}\left[ (1+t^{\alpha})^{\theta}-1\right]^{k}e^{-\lambda[(1+t^{\alpha})^{\theta}-1]}}{\sum_{k=0}^{s}a_{k}\frac{\Gamma\left(k+1, \lambda[(1+t^{\alpha})^{\theta}-1]\right)}{\lambda^{k+1}}}.
\end{eqnarray*}
\textbf{Odds Lindley - Burr XII Distribution:-}\\
In equation number $(\ref{eq2})$, when $s=1$, $a_0=1$, $a_1=1$, the One Parameter Polynomial Exponential gives the Lindley distribution. So when $s=1$, $a_0=1$, $a_1=1$, the Odds OPPE - Burr XII Distribution reduces to Odds Lindley - Burr XII Distribution with pdf 
\begin{eqnarray*}
f(x;\lambda,\alpha,\theta)&=&\frac{\lambda^2}{1+\lambda}\alpha\theta x^{\alpha-1}\left(1+x^\alpha\right)^{2\theta-1}e^{-\lambda[(1+x^{\alpha})^{\theta}-1]}; \nonumber\\&& 0<x,\theta,\alpha<\infty, \lambda>0.
\end{eqnarray*}
\begin{figure*}
  \includegraphics[width=1\textwidth]{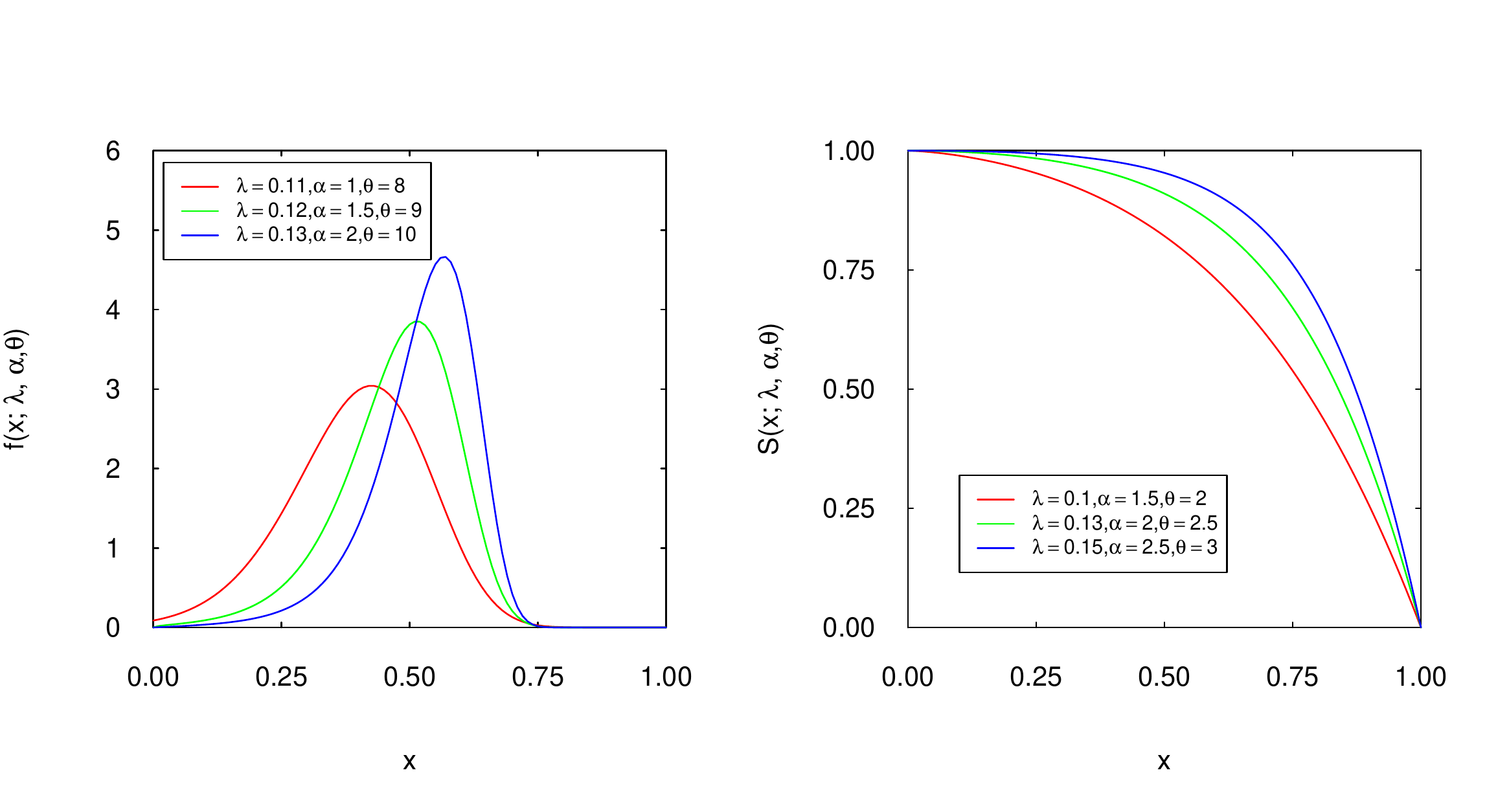}
\caption{The pdf and survival function of Odds Lindley - Bur XII distribution}
\label{fig4}
\end{figure*}

\section{Some Mathematical Properties}
In this section, some general results of the Odds OPPE - G family are derived.
\subsection{Mathematical Expansions}
Expansion formulae of the Odds OPPE - G family, such as; the pdf and cdf are derived. The probability density function (pdf) of Odds OPPE - G family of distribution, is given by
\begin{eqnarray}\label{eq7}
f(x;\lambda,\xi)&=&h(\lambda)\sum_{k=0}^{s}a_{k}\frac{g(x;\xi)}{[\bar{G}(x;\xi)]^2}\left[\frac{G(x;\xi)}{\bar{G}(x;\xi)}\right]^{k}e^{-\lambda \left[\frac{G(x;\xi)}{\bar{G}(x;\xi)}\right]}\nonumber\\&=&h(\lambda)\sum_{k=0}^{s}a_{k}\frac{g(x;\xi)}{[\bar{G}(x;\xi)]^2}\left[\frac{G(x;\xi)}{\bar{G}(x;\xi)}\right]^{k}\sum_{i=0}^\infty\frac{(-1)^i}{i!}{\lambda^i}\left[\frac{G(x;\xi)}{\bar{G}(x;\xi)}\right]^i\nonumber\\&=&h(\lambda)\sum_{k=0}^{s}\sum_{i=0}^\infty\frac{(-1)^{i}a_{k}\lambda^{i}}{i!}\frac{g(x;\xi)[G(x;\xi)]^{k+i}}{[\bar{G}(x;\xi)]^{k+i+2}}\nonumber\\&=&h(\lambda)\sum_{k=0}^{s}\sum_{i=0}^\infty\frac{(-1)^{i}a_{k}\lambda^{i}}{i!}g(x;\xi)[G(x;\xi)]^{k+i}[\bar{G}(x;\xi)]^{-(k+i+2)}\nonumber\\&=&h(\lambda)\sum_{k=0}^{s}\sum_{i,j=0}^\infty\frac{(-1)^{i}a_{k}\lambda^{i}}{i!}\binom{i+j+k+1}{j}g(x;\xi)[G(x;\xi)]^{i+j+k}\nonumber\\&=&\frac{\sum_{k=0}^{s}\sum_{i,j=0}^\infty w_{ijk}(\lambda) h_{i+j+k}(x;\xi)}{\sum_{k=0}^{s}w_{k}(\lambda)}
\end{eqnarray}
where, $w_{ijk}(\lambda)=\sum_{k=0}^{s}\sum_{i,j=0}^\infty\frac{(-1)^{i}a_{k}\lambda^{i}}{i!}\binom{i+j+k+1}{j}$, $w_{k}(\lambda)=a_{k}\frac{\Gamma(k+1)}{\lambda^{k+1}}$, and \\ $h_{i+j+k}(x;\xi)=g(x;\xi)[G(x;\xi)]^{i+j+k}$.\\
The cdf of X is given by 
\begin{eqnarray}\label{eq8}
F(x;\lambda,\xi)&=&\int_{0}^{x}f(t;\lambda,\xi)dt\nonumber\\&=&\frac{\sum_{k=0}^{s}\sum_{i,j=0}^\infty w_{ijk}(\lambda)\int_{0}^{x}g(t;\xi)[G(t;\xi)]^{i+j+k}dt}{\sum_{k=0}^{s}w_{k}(\lambda)}\nonumber\\&=&\frac{\sum_{k=0}^{s}\sum_{i,j=0}^\infty w_{ijk}(\lambda)\frac{[G(x;\xi)]^{i+j+k+1}}{i+j+k+1}}{\sum_{k=0}^{s}w_{k}(\lambda)}.
\end{eqnarray}

\subsection{Shapes of the Odds OPPE - G family of distribution}
The shapes of the density and hazard rate functions can also be described analytically.\\
Now, 
\begin{eqnarray*}
	f(x;\lambda,\xi)&=&h(\lambda)\sum_{k=0}^{s}a_{k}\frac{g(x;\xi)}{[\bar{G}(x;\xi)]^2}\left[\frac{G(x;\xi)}{\bar{G}(x;\xi)}\right]^{k}e^{-\lambda \left[\frac{G(x;\xi)}{\bar{G}(x;\xi)}\right]}.
\end{eqnarray*}
where, $h(\lambda)=\frac{1}{\sum_{k=0}^{s}a_{k}\frac{\Gamma(k+1)}{\lambda^{k+1}}}$.\\
So, \begin{eqnarray*}	\ln f(x;\lambda,\xi)&=& \ln h(\lambda) + \ln \sum_{k=0}^{s}a_{k}\frac{g(x;\xi)}{[\bar{G}(x;\xi)]^2}\left[\frac{G(x;\xi)}{\bar{G}(x;\xi)}\right]^{k}e^{-\lambda \left[\frac{G(x;\xi)}{\bar{G}(x;\xi)}\right]}.
\end{eqnarray*}
Now, the critical points of the Odds OPPE - G density function are the roots of the equation:
\begin{eqnarray*}\frac{d}{dx}\ln f(x;\lambda,\xi)&=&\frac{d}{dx}\ln \left\lbrace \sum_{k=0}^{s}a_{k}\frac{g(x;\xi)}{[\bar{G}(x;\xi)]^2}\left[\frac{G(x;\xi)}{\bar{G}(x;\xi)}\right]^{k}e^{-\lambda \left[\frac{G(x;\xi)}{\bar{G}(x;\xi)}\right]}\right\rbrace =0.
\end{eqnarray*}

\subsection{Quantile function}
The quantile function, say $Q(u)=F^{-1}(u)$,  of the Odds OPPE - G family is derived by inverting $(\ref{eq3})$ as follows
\begin{eqnarray*}
	u&=&1-\frac{\sum_{k=0}^{s}a_{k}\frac{\Gamma\left(k+1, \lambda \frac{Q(u)}{1-Q(u)}\right)}{\lambda^{k+1}}}{\sum_{k=0}^{s}a_{k}\frac{\Gamma(k+1)}{\lambda^{k+1}}}.
\end{eqnarray*}
So,
\begin{eqnarray*}
	\sum_{k=0}^{s}a_{k}\frac{\Gamma\left(k+1, \lambda \frac{Q(u)}{1-Q(u)}\right)}{\lambda^{k+1}}&=&(1-u)\sum_{k=0}^{s}a_{k}\frac{\Gamma(k+1)}{\lambda^{k+1}}.
\end{eqnarray*}
Taking Logarithm on both sides, the previous equation is reduced to
\begin{eqnarray}\label{eq9}
\ln\sum_{k=0}^{s}a_{k}\frac{\Gamma\left(k+1, \lambda \frac{Q(u)}{1-Q(u)}\right)}{\lambda^{k+1}}-\ln(1-u)-\ln\sum_{k=0}^{s}a_{k}\frac{\Gamma(k+1)}{\lambda^{k+1}}=0.
\end{eqnarray}
By solving the nonlinear equation $(\ref{eq9})$, numerically, the Odds OPPE - G family random variable X can be generated, where u has the uniform distribution on the unit interval.

\subsection{Moments}
The $r^{th}$ moment of random variable X can be obtained from pdf $(\ref{eq7})$ as follows
\begin{eqnarray*}
	\mu_r^{'}&=&\int_0^{\infty}x^{r}f(x,\lambda,\xi)dx\nonumber\\&=&\frac{\sum_{k=0}^{s}\sum_{i,j=0}^\infty w_{ijk}(\lambda)\int_{0}^{\infty}x^{r}h_{i+j+k}(x;\xi)dx}{\sum_{k=0}^{s}w_{k}(\lambda)}.
\end{eqnarray*}
Therefore,
\begin{eqnarray}\label{eq10}
	\mu_r^{'}&=& \frac{\sum_{k=0}^{s}\sum_{i,j=0}^\infty w_{ijk}(\lambda)I_{i,j,k,r}}{\sum_{k=0}^{s}w_{k}(\lambda)};~r=1,2,....
\end{eqnarray}
where,
$I_{i,j,k,r}=\int_0^{\infty}x^{r} h_{i+j+k}(x;\xi)dx$.\\
In particular, the mean and variance of Odds OPPE - G family are obtained as follows:
\begin{eqnarray*}
E(X)&=& \frac{\sum_{k=0}^{s}\sum_{i,j=0}^\infty w_{ijk}(\lambda)I_{i,j,k,1}}{\sum_{k=0}^{s}w_{k}(\lambda)},
\end{eqnarray*}

\begin{eqnarray*}
Var(X)&=&\frac{\sum_{k=0}^{s}\sum_{i,j=0}^\infty w_{ijk}(\lambda)I_{i,j,k,2}}{\sum_{k=0}^{s}w_{k}(\lambda)}-\left[\frac{\sum_{k=0}^{s}\sum_{i,j=0}^\infty w_{ijk}(\lambda)I_{i,j,k,1}}{\sum_{k=0}^{s}w_{k}(\lambda)}\right]^2.
\end{eqnarray*}
Additionally, measures of skewness and kurtosis of the family can be obtained, based on $(\ref{eq10})$, according to the following relations
\begin{eqnarray*}
	\gamma_{1}&=&\frac{\mu_3^{'}-3\mu_2^{'}\mu_1^{'}+2\mu_1^{'^3}}{\left(\mu_2^{'}-\mu_1^{'^2}\right)^{3/2}},
\end{eqnarray*}
\begin{eqnarray*}
	\gamma_{2}&=&\frac{\mu_4^{'}-4\mu_3^{'}\mu_1^{'}+6\mu_2^{'}\mu_1^{'^2}-3\mu_1^{'^4}}{\left(\mu_2^{'}-\mu_1^{'^2}\right)^2}.
\end{eqnarray*}

\subsection{Generating Function}
\textbf{The Moment Generating function(MGF)} of Odds OPPE - G family is defined as
\begin{eqnarray*}
M_{X}(t)&=&\sum_{r=0}^\infty \frac{t^{r}}{r!} \mu_{r}^{'},
\end{eqnarray*}
where, $\mu_r^{'}$  is the $r^{th}$ moment about origin. Then  the moment generating function of Odds OPPE - G family is obtained by using $(\ref{eq10})$ as follows
\begin{eqnarray*}
M_{X}(t)&=&\sum_{r=0}^\infty \frac{t^{r}}{r!} \left[ \frac{\sum_{k=0}^{s}\sum_{i,j,r=0}^\infty w_{ijk}(\lambda)I_{i,j,k,r}}{\sum_{k=0}^{s}w_{k}(\lambda)}\right].
\end{eqnarray*}

\textbf{Characteristic Function(CF)}: \begin{eqnarray*} \Psi_{X}(t)&=& E(e^{itX})\nonumber\\&=&\sum_{r=0}^\infty \frac{(it)^{r}}{r!} \mu_{r}^{'}\nonumber\\&=&\sum_{r=0}^\infty\frac{(it)^{r}}{r!}\left[ \frac{\sum_{k=0}^{s}\sum_{i,j,r=0}^\infty w_{ijk}(\lambda)I_{i,j,k,r}}{\sum_{k=0}^{s}w_{k}(\lambda)}\right].\end{eqnarray*}

\textbf{Cumulant Generating Function(CGF)}: \begin{eqnarray*} K_{X}(t)&=&\ln(M_{X}(t))\nonumber\\&=&\ln\sum_{r=0}^\infty \frac{t^{r}}{r!} \left[ \frac{\sum_{k=0}^{s}\sum_{i,j,r=0}^\infty w_{ijk}(\lambda)I_{i,j,k,r}}{\sum_{k=0}^{s}w_{k}(\lambda)}\right].\end{eqnarray*}

\subsection{Entropy}
The variation of the uncertainty in a random X is sometimes measured by entropy. It is specially used for random variables having heavy-tail distribution when all or some order moments are non-existent. Renyi[29] entropy is a more general entropy measure. For a random variable X having a probability density function f(x), the Renyi entropy for the Odds OPPE-G distribution is defined by 
\begin{eqnarray}\label{eq11}
H_R(\beta)&=&\frac{1}{1-\beta}\ln\left\{\int_{0}^{\infty}f^\beta(x)dx \right\}\nonumber\\&=&\frac{1}{1-\beta}\ln\left\{\frac{\int_{0}^{\infty}\left[ \sum_{k=0}^{s}\sum_{i,j=0}^\infty w_{ijk}(\lambda) h_{i+j+k}(x;\xi)\right]^{\beta} dx}{\left[ \sum_{k=0}^{s}w_{k}(\lambda)\right]^{\beta}} \right\},
\end{eqnarray}
where $\beta>0$, $\beta \neq 1$, $w_{ijk}(\lambda)=\sum_{k=0}^{s}\sum_{i,j=0}^\infty\frac{(-1)^{i}a_{k}\lambda^{i}}{i!}\binom{i+j+k+1}{j}$, $w_{k}(\lambda)=a_{k}\frac{\Gamma(k+1)}{\lambda^{k+1}}$, and $h_{i+j+k}(x;\xi)=g(x;\xi)[G(x;\xi)]^{i+j+k}$.\\
The Shannon entropy is given by $E\left\lbrace -log[f(x)]\right\rbrace $. It is a special case of the Renyi entropy
when $\beta \rightarrow 1$.

{\bf Example 3.1:-}\\
{\it Consider the Odds Lindley - Exponential distribution discussed in subsection 2.2.}\\
Renyi entropy for Odds Lindley - Exponential distribution is 
\begin{eqnarray*}H_R(\beta)&=&-\ln \theta +\frac{\lambda \beta}{1-\beta}-\frac{\beta}{1-\beta}\ln(1+\lambda)-\frac{2\beta}{1-\beta}\ln \beta +\frac{\ln \Gamma\left(2\beta,\lambda\beta\right)}{1-\beta}\end{eqnarray*}

Shannon measure of entropy for Odds Lindley - Exponential distribution 
\begin{eqnarray*}
H(f)=E[-\ln f(x)]&=&-2\ln \lambda -\ln \theta - \lambda +\ln(1+\lambda)+\frac{e^{\lambda}}{1+\lambda}\Gamma\left(3,\lambda\right)\nonumber\\&&-\frac{2 e^{\lambda}}{(1+\lambda)}\left[\Gamma^{(1)}(2,\lambda)-\ln\lambda.\Gamma(2,\lambda)\right]\end{eqnarray*}

{\bf Example 3.2:-}\\
{\it Consider the Odds Lindley - Pareto distribution discussed in subsection 2.3.}\\
Renyi entropy for Odds Lindley - Pareto distribution is \begin{eqnarray*}H_R(\beta)&=&-\frac{\ln\lambda}{\theta}-\ln\theta +\ln a+ \frac{\lambda\beta}{1-\beta} - \frac{(2\beta-\frac{\beta}{\theta}+\frac{1}{\theta})}{1-\beta}\ln\beta +\frac{1}{1-\beta}\ln\Gamma\left(2\beta-\frac{\beta}{\theta}+\frac{1}{\theta},\lambda\beta\right)\nonumber\\&&-\frac{\beta}{1-\beta}\ln(1+\lambda)\end{eqnarray*}

Shannon measure of entropy for Odds Lindley - Pareto distribution is \begin{eqnarray*}
H(f)=E[-\ln f(x)]&=&-\lambda-\ln \theta+\ln a-\frac{\ln \lambda}{\theta}+\ln(1+\lambda)+\frac{e^{\lambda}}{1+\lambda}\Gamma\left(3,\lambda\right)-\frac{(2\theta-1)e^{\lambda}}{(1+\lambda)\theta}\Gamma^{(1)}(2,\lambda)\end{eqnarray*}

\subsection{Order Statistics}
A branch of statistics known as order statistics plays a prominent role in real-life applications involving data relating to life testing studies. These statistics are required in many fields, such as climatology, engineering and industry, among others. A comprehensive exposition of order statistics and associated inference is provided by David and Nagaraja[10]. Let $X_{r:n}$ denote the $r^{th}$ order statistic. The density $f_{r:n}(x)$ of the $r^{th}$ order statistic, for $r = 1(1)n,$ from independent and identically distributed random variables $X_1, X_2,.....  X_n$ having the Odds OPPE-G distribution is given by
\begin{eqnarray*}
f_{r:n}(x)&=&M.\left[F(x)\right]^{r-1}\left[1-F(x)\right]^{n-r}f(x)\nonumber\\&=&M.\sum_{l=0}^{n-r}(-1)^l\binom{n-r}{l}\left[F(x)\right]^{r+l-1}f(x),
\end{eqnarray*}
where $M=\frac{n!}{(r-1)!(n-r)!}$\\
So, \begin{eqnarray}\label{eq11}
f_{r:n}(x;\Phi)&=& M.\sum_{l=0}^{n-r}(-1)^l\binom{n-r}{l}\left[1-h(\lambda)\sum_{k=0}^{s}a_{k}\frac{\Gamma\left(k+1, \lambda \frac{G(x;\xi)}{\bar{G}(x;\xi)}\right)}{\lambda^{k+1}}\right]^{r+l-1}\nonumber\\&&.h(\lambda)\left[ \sum_{k=0}^{s}a_{k}\frac{g(x;\xi)}{[\bar{G}(x;\xi)]^2}\left[\frac{G(x;\xi)}{\bar{G}(x;\xi)}\right]^{k}e^{-\lambda \left[\frac{G(x;\xi)}{\bar{G}(x;\xi)}\right]}\right].
\end{eqnarray}
where, $h(\lambda)=\frac{1}{\sum_{k=0}^{s}a_{k}\frac{\Gamma(k+1)}{\lambda^{k+1}}}$.

\subsection{Stress-Strength Reliability}
The measure of reliability of industrial components has many applications especially in the area of engineering. The reliability of a product (system) is the probability that the product (system) will perform its intended function for a specified time period when operating under normal (or stated) environmental conditions. The component fails at the instant that the random stress $X_2$ applied to it exceeds the random strength $X_1$, and the component will function satisfactorily whenever $X_1 > X_2$. Hence, $R = P(X_2 < X_1)$ is a measure of component reliability [see Kotz, Lai, and Xie[17]]. We derive the reliability R when $X_1$ and $X_2$ have independent Odds OPPE-G(x; $\lambda_1;\xi$) and Odds OPPE-G(x; $\lambda_2;\xi$) distributions with the same parameter vector $\xi$ for the baseline G. The reliability is denoted by
\begin{eqnarray*} R \nonumber&=&\int_{0}^{\infty}f_{1}(x)F_{2}(x)dx\end{eqnarray*}
The pdf of $X_1$ and cdf of $X_2$ are obtained from equation $(\ref{eq7})$ and $(\ref{eq8})$ as 
\begin{eqnarray*}
f_{1}(x)&=&\frac{\sum_{k=0}^{s}\sum_{i,j=0}^\infty w_{ijk}(\lambda_1) g(x;\xi)[G(x;\xi)]^{i+j+k}}{\sum_{k=0}^{s}w_{k}(\lambda_1)}
\end{eqnarray*}
\begin{eqnarray*}
F_{2}(x)&=&\frac{\sum_{k=0}^{s}\sum_{i,j=0}^\infty w_{ijk}(\lambda_2)\frac{G(x;\xi)]^{i+j+k+1}}{i+j+k+1}}{\sum_{k=0}^{s}w_{k}(\lambda_2)}
\end{eqnarray*}
where, $w_{ijk}(\lambda_1)=\sum_{k=0}^{s}\sum_{i,j=0}^\infty\frac{(-1)^{i}a_{k}\lambda_{1}^{i}}{i!}\binom{i+j+k+1}{j}$, $w_{k}(\lambda_1)=a_{k}\frac{\Gamma(k+1)}{\lambda_{1}^{k+1}}$, \\$w_{ijk}(\lambda_2)=\sum_{k=0}^{s}\sum_{i,j=0}^\infty\frac{(-1)^{i}a_{k}\lambda_{2}^{i}}{i!}\binom{i+j+k+1}{j}$, and $w_{k}(\lambda_2)=a_{k}\frac{\Gamma(k+1)}{\lambda_{2}^{k+1}}$.\\
Hence, \begin{eqnarray*} R \nonumber&=&\int_{0}^{\infty}f_{1}(x)F_{2}(x)dx\nonumber\\&=&\int_{0}^{\infty}\left\lbrace \frac{\sum_{k=0}^{s}\sum_{i,j=0}^\infty w_{ijk}(\lambda_1) g(x;\xi)[G(x;\xi)]^{i+j+k}}{\sum_{k=0}^{s}w_{k}(\lambda_1)}.\frac{\sum_{k=0}^{s}\sum_{i,j=0}^\infty w_{ijk}(\lambda_2)\frac{[G(x;\xi)]^{i+j+k+1}}{i+j+k+1}}{\sum_{k=0}^{s}w_{k}(\lambda_2)}\right\rbrace dx\end{eqnarray*}

\begin{figure*}
  \includegraphics[width=.75\textwidth]{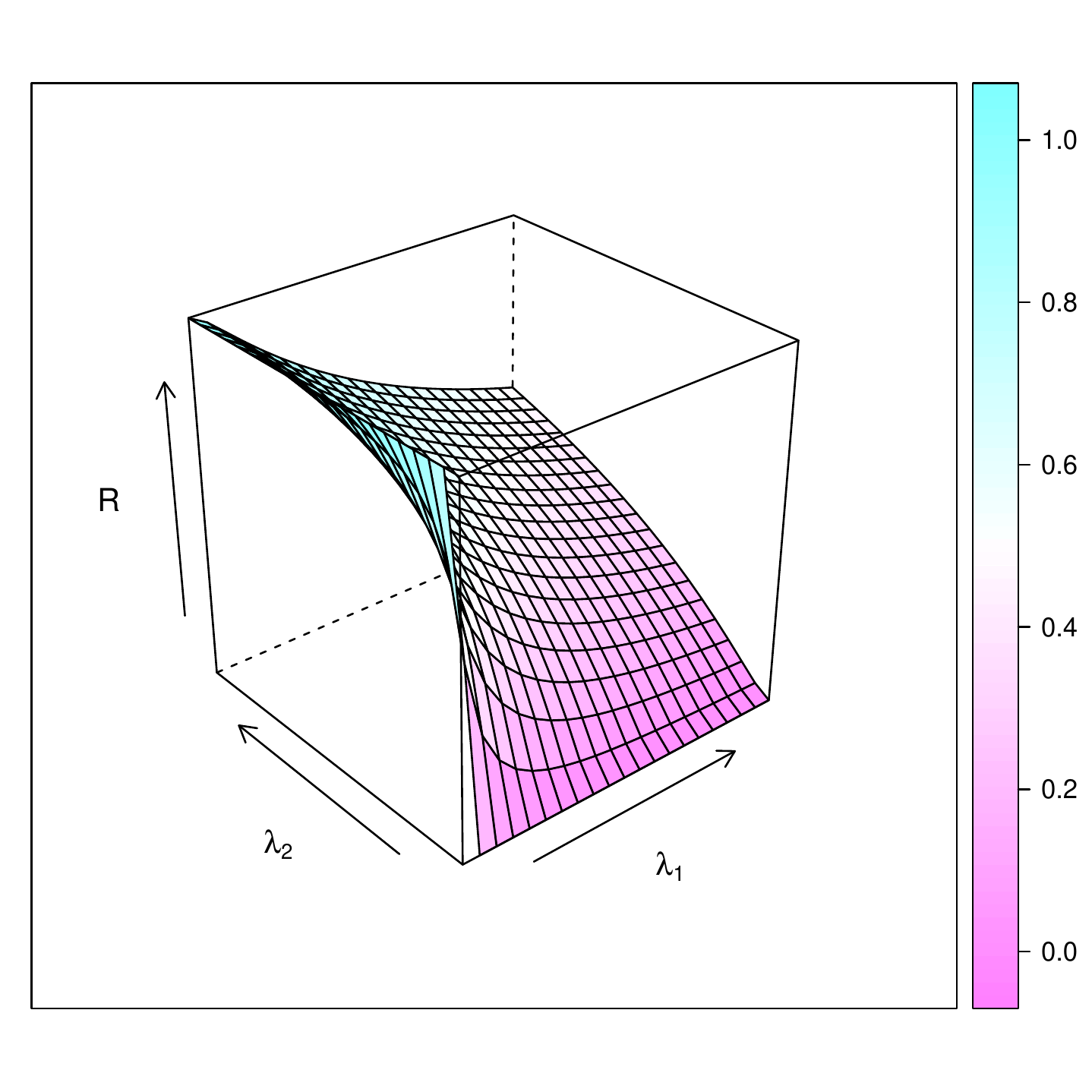}
\caption{Stress-Strength Reliability, R for different $\lambda_1$ and $\lambda_2$ when $\theta_1=\theta_2$, of Odds Lindley Exponential distribution}
\label{fig5}
\end{figure*}

\begin{figure*}
  \includegraphics[width=.75\textwidth]{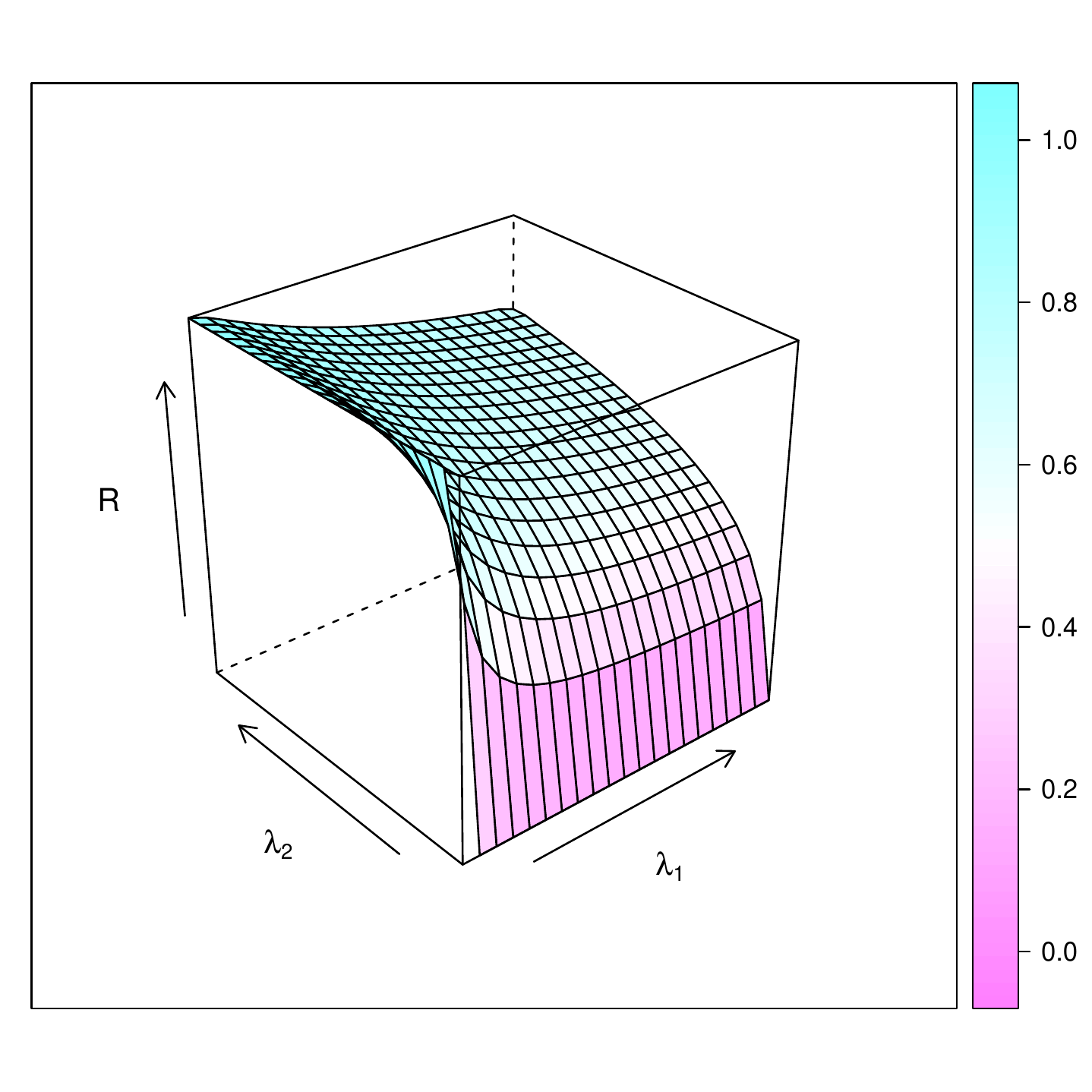}
\caption{Stress-Strength Reliability, R for different $\lambda_1$ and $\lambda_2$ when $\theta_1=\theta_2$, and $a_1=a_2$ of Odds Lindley Pareto distribution}
\label{fig6}
\end{figure*}

{\bf Example 3.3:-}\\
{\it Consider the Odds Lindley - Exponential distribution discussed in subsection 2.2.} \\
Let $X_1 \sim OLED(\lambda_{1}, \theta_{1})$ and $X_2 \sim OLED(\lambda_{2}, \theta_{2})$ be independent random variables. Then Stress-Strength Reliability \begin{eqnarray*} R \nonumber&=&P(X_2 < X_1) \nonumber\\&=&1- \frac{{\lambda_1}{^2}\theta_1 e^{\lambda_1+\lambda_2}}{(1+\lambda_1)(1+\lambda_2)}\int_{0}^{\infty}[1+ \lambda_2 e^{\theta_{2}x}]e^{2\theta_{1}x}e^{-\lambda_1 e^{\theta_{1}x}-\lambda_2 e^{\theta_{2}x}}dx\end{eqnarray*}
If $\theta_1=\theta_2=\theta$,
\begin{eqnarray*}
R \nonumber&=&1- \frac{{\lambda_1}^{2}}{(1+\lambda_1)(1+\lambda_2)}\left[\frac{1+ \lambda_2}{\lambda_1+ \lambda_2}+\frac{1+ 2\lambda_2}{(\lambda_1+ \lambda_2)^2}+\frac{2\lambda_2}{(\lambda_1+ \lambda_2)^3}\right]\end{eqnarray*}

{\bf Example 3.4:-}\\
{\it Consider the Odds Lindley - Pareto distribution discussed in subsection 2.3.} \\
Let $X_1 \sim OLPD(\lambda_{1}, \theta_{1}, a_{1})$ and $X_2 \sim OLPD(\lambda_{2}, \theta_{2}, a_{2})$ be independent random variables. Then Stress-Strength Reliability \begin{eqnarray*} R \nonumber&=&P(X_2 < X_1) \nonumber\\&=&1- \frac{{\lambda_1}{^2}\theta_1 e^{\lambda_1+\lambda_2}}{(1+\lambda_1)(1+\lambda_2)a_1^{2\theta_1}}\int_{a_{1}}^{\infty}[1+ \lambda_2 (\frac{x}{a_2})^{\theta_2}]x^{2\theta_1-1}e^{-\lambda_1(\frac{x}{a_1})^{\theta_1}-\lambda_2(\frac{x}{a_2})^{\theta_2}}dx\end{eqnarray*}
If $\theta_1=\theta_2=\theta$, then \begin{eqnarray*} R \nonumber&=&1- \frac{{\lambda_1}^{2}e^{\lambda_1+\lambda_2}}{(1+\lambda_1)(1+\lambda_2) a_1^{2\theta}}\left[\frac{\Gamma(2,\lambda_1+\lambda_2\frac{ {a_1}^\theta}{{a_2}^\theta})}{(\frac{\lambda_1}{a_1^\theta}+\frac{\lambda_2}{a_2^\theta})^2}+\frac{\lambda_2}{a_2^\theta}\frac{\Gamma(3,\lambda_1+\lambda_2\frac{ {a_1}^\theta}{{a_2}^\theta})}{(\frac{\lambda_1}{a_1^\theta}+\frac{\lambda_2}{a_2^\theta})^3}\right]\end{eqnarray*}
Also if $a_1=a_2$, then \begin{eqnarray*} R \nonumber&=&1- \frac{{\lambda_1}^{2}e^{\lambda_1+\lambda_2}}{(1+\lambda_1)(1+\lambda_2) (\lambda_1+\lambda_2)^2}\left[\Gamma(2,\lambda_1+\lambda_2)+\frac{\lambda_2}{\lambda_1+\lambda_2}\Gamma(3,\lambda_1+\lambda_2)\right]\end{eqnarray*}

\subsection{Incomplete Moments, Mean Deviations, and Lorenz and Benferroni Curves}
The $r^{th}$ incomplete moment, say, $m_r^I(t)$, of the Odds OPPE - G Family of distributions is given by
\begin{eqnarray*}
m_r^I(t)&=& \int_0^tx^rf(x,\Phi)dx.
\end{eqnarray*}
We can write from equation ($\ref{eq7}$),
\begin{eqnarray}\label{eq12}
m_r^I(t)&=& \int_0^tx^r\left[\frac{\sum_{k=0}^{s}\sum_{i,j=0}^\infty w_{ijk}(\lambda) g(x;\xi)[G(x;\xi)]^{i+j+k}}{\sum_{k=0}^{s}w_{k}(\lambda)}\right]dx.
\end{eqnarray}

{\bf Example 3.5:-}
\\{\it $r^{th}$ incomplete moment for Odds Lindley - Exponential distribution is}
\begin{eqnarray*}
m_r^I(t)&=&\int_{0}^{t}x^rf(x)dx\\&=& \frac{e^{\lambda}}{(1+\lambda)\theta^r}\sum_{j=0}^r(-1)^{r-j}\binom{r}{j}\left(\ln{\lambda}\right)^{r-j}\left\lbrace \Gamma^{(j)}(2, \lambda)-\Gamma^{(j)}(2, \lambda e^\theta)\right\rbrace .
\end{eqnarray*}
{\it $r^{th}$ incomplete moment for Odds Lindley - Pareto distribution is}
\begin{eqnarray*} m_r^I(t)&=&\int_{a}^{t}x^rf(x)dx\\&=& \frac{e^{\lambda}a^r}{(1+\lambda)\lambda^{\frac{r}{\theta}}}\left[\Gamma\left(\frac{r}{\theta}+2,\lambda\right)-\Gamma\left(\frac{r}{\theta}+2,\lambda \left(\frac{t}{a}\right)^{\theta}\right)\right]
\end{eqnarray*} 

Apart from range and s.d., mean deviation about mean, $\delta_1$ and median, $\delta_2$ are used as measures of spread in a population. Incomplete moments are used to define $\delta_1=2\mu_1^{'}F(\mu_1^{'})-2m_1^I(\mu_1^{'})$ and $\delta_2=\mu_1^{'}-2m_1^I(\mu_e)$,respectively. Here, $\mu_1^{'}=E(X)$ is to be obtained from ($\ref{eq9}$) with $r=1$, $F(\mu_1^{'})$ is to calculated from ($\ref{eq2}$), $m_1^I(\mu_1^{'})$ is the first incomplete function obtained from ($\ref{eq12}$) with $r=1$ and $\mu_e$ is the median of $X$ obtained by solving ($\ref{eq8}$) for $u=0.5$.\\

The Lorenz and Benferroni curves are defined by $L(p)=m_1^I(x_p)/\mu_1^{'}$ and $B(p)=m_1^I(x_p)/(p\mu_1^{'})$, respectively, where $x_p=F^{-1}(p)$ can be computed numerically by ($\ref{eq8}$) with $u=p$. These curves are significantly used in economics, reliability, demography, insurance and medicine. For details in this aspect, we refer to Pundir, Arora, and Jain[28] and references cited therein.

\begin{figure*}
  \includegraphics[width=1\textwidth]{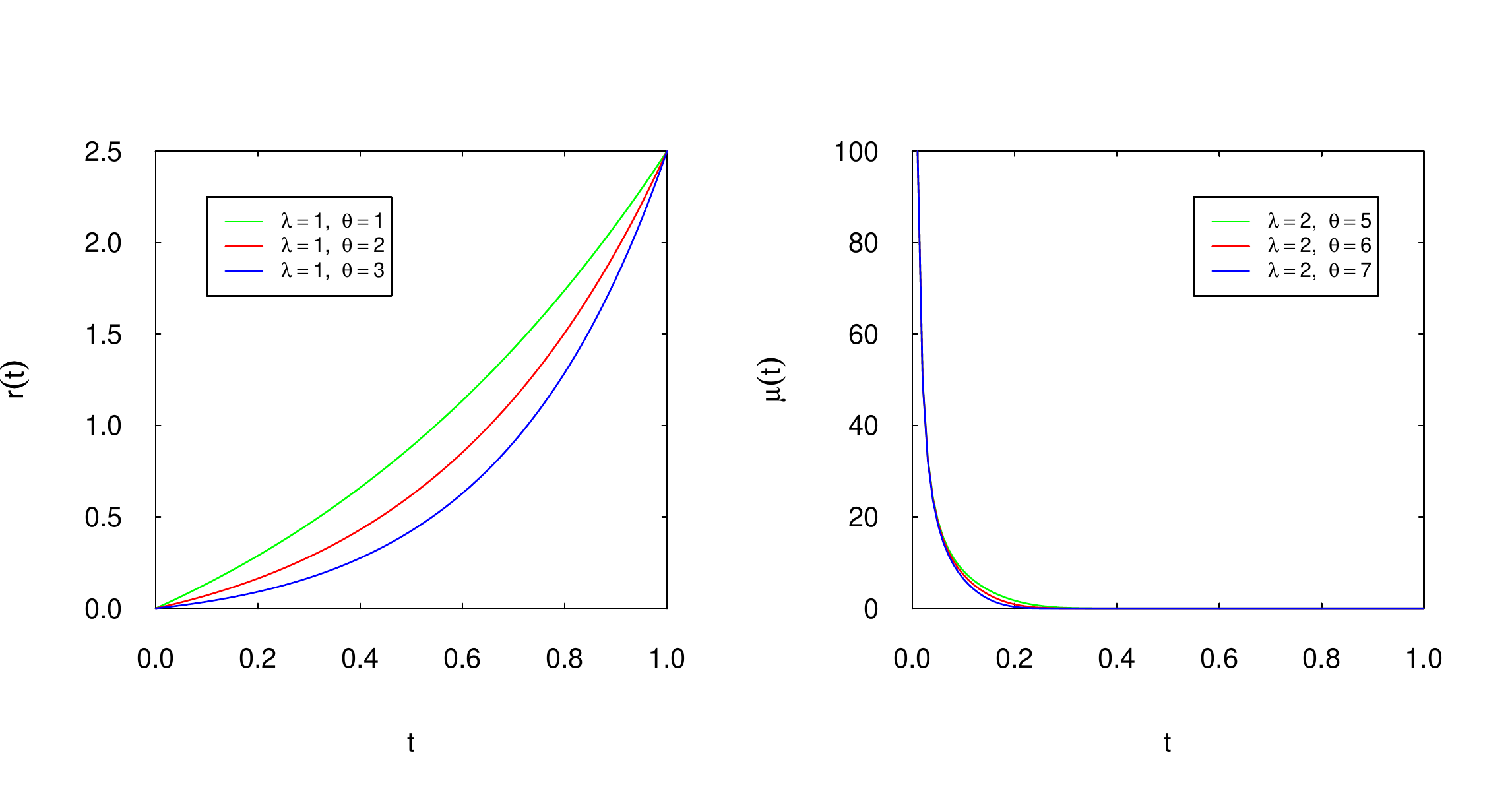}
\caption{The Hazard Rate and Reversed Hazard Rate of Odds Lindley - Exponential distribution}
\label{fig7}
\end{figure*}

\begin{figure*}
  \includegraphics[width=1\textwidth]{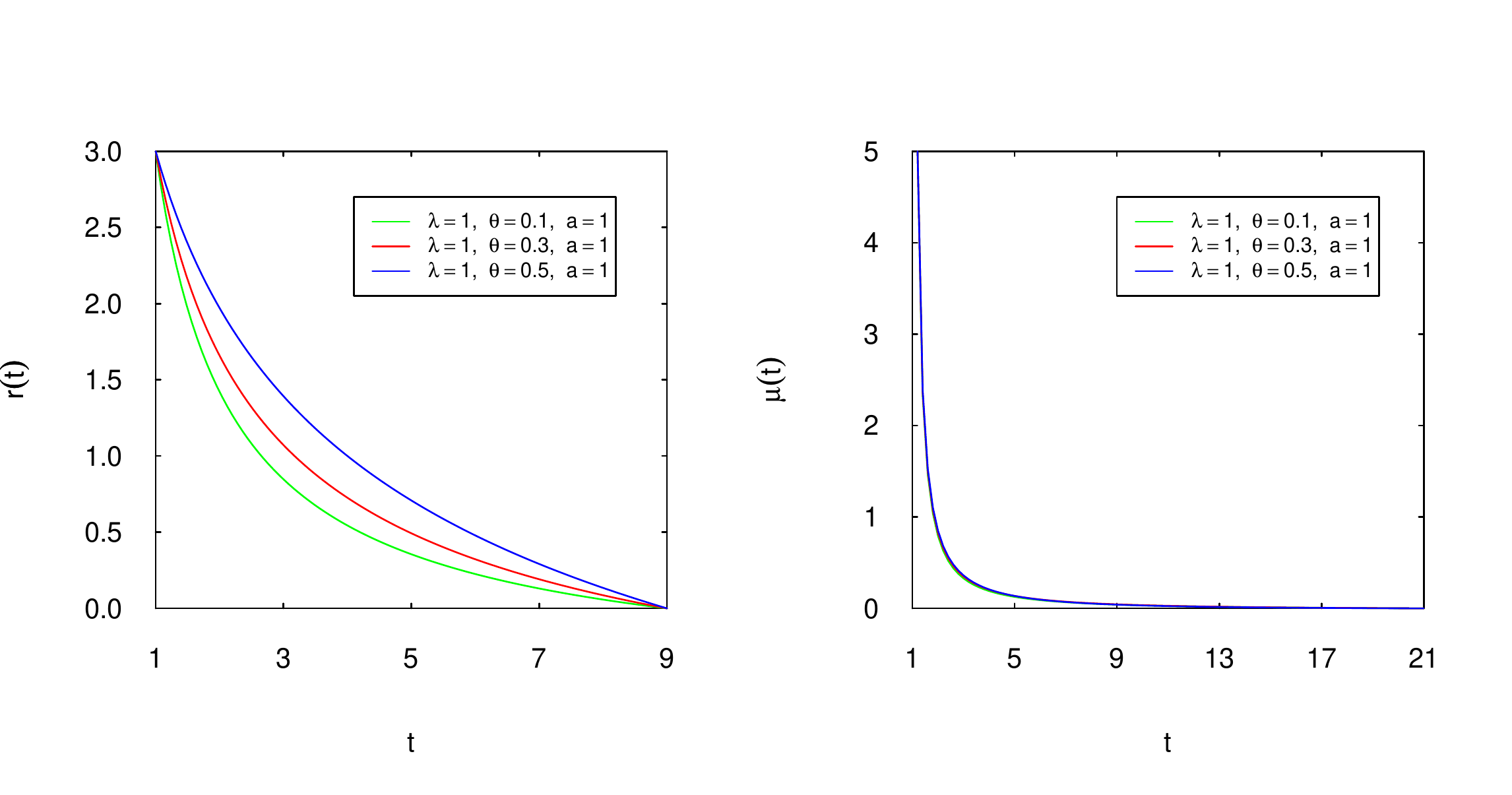}
\caption{The Hazard Rate and Reversed Hazard Rate of Odds Lindley - Pareto distribution}
\label{fig8}
\end{figure*}

\subsection{Moments of the residual life}
The residual life function plays an important role in reliability/survival analysis, social studies, bio-medical sciences, economics, population study, insurance industry, maintenance and product quality control and product technology. If $X$ be a random variable denoting the lifetime of a unit at age $t$, then $X_t=X-t\mid X>t$ is the remaining lifetime beyond that age $t$.

The cdf $F(x)$ is uniquely determined by the $r^{th}$ moment of the residual life of $X$ (for $r=1,~2,~...$) [Navarro, Franco, and Ruiz[26]], and it is given by
\begin{eqnarray*}
	m_r(t)=E[X_t]&=& \frac{1}{\bar F(t)}\int_t^{\infty}(x-t)^rdF(x)\\&=&\frac{1}{1-F(t)}\int_t^{\infty}(x-t)^{r}f(x,\Phi)dx.
\end{eqnarray*}
In particular, if $r=1$, then $m_1(t)$ represents an interesting function, called the mean residual life (MRL) function that represents the average life length for a unit which is alive at age $t$.\\

{\bf Example 3.6:-}\\
{\it Consider the Odds Lindley - Exponential distribution discussed in subsection 2.2.} 
\begin{eqnarray*}
m_r(t)&=&\frac{e^{\lambda e^{\theta t}}}{1+ \lambda e^{\theta t}}\sum_{j=0}^r \frac{(-1)^j}{\theta ^j}{{r}\choose j}t^{r-j}\sum_{k=0}^j(-1)^{j-k}\binom{j}{k}\left(\ln{\lambda}\right)^{j-k}\Gamma^{(k)}(2, \lambda e^{\theta t})
\end{eqnarray*}
For the MRL function,
\begin{eqnarray*}
m_1(t)&=&\frac{e^{\lambda e^{\theta t}}}{1+\lambda e^{\theta t}}\left[\frac{1}{\theta}\Gamma^{(1)}(2,\lambda e^{\theta t})-\left(t+\frac{\ln\lambda}{\theta}\right)\Gamma(2,\lambda e^{\theta t})\right].
\end{eqnarray*}

{\bf Example 3.7:-}\\
{\it Consider the Odds Lindley - Pareto distribution discussed in subsection 2.3.} 
\begin{eqnarray*}
m_r(t)&=&\frac{e^{\lambda(\frac{t}{a})^\theta}}{1+ \lambda (\frac{t}{a})^\theta}\sum_{j=0}^r (-1)^j {{r}\choose j}t^{r-j}\frac{a^j}{\lambda^\frac{j}{\theta}}\Gamma(\frac{j}{\theta}+2,\frac{\lambda t^\theta}{a^\theta})
\end{eqnarray*}
For the MRL function,
\begin{eqnarray*}
m_1(t)&=&\frac{e^{\lambda(\frac{t}{a})^\theta}}{1+ \lambda (\frac{t}{a})^\theta}\left[\frac{a}{\lambda^\frac{1}{\theta}}\Gamma(2+\frac{1}{\theta},\frac{\lambda t^\theta}{a^\theta})-t\Gamma(2,\frac{\lambda t^\theta}{a^\theta})\right].
\end{eqnarray*}

\subsection{Moments of the reversed residual life}
Some real life situations are there where uncertainty is not only related to the future but can also refer to the past. Consider a system whose state is observed only at certain preassigned inspection time $t$. If the system is inspected for the first time and it is found to be `down', then failure relies on the past i.e. on which instant in $(0,t)$ it has failed. So, study of a dual notion to the residual life that deal with the past time seems worthwhile [see Di Crescenzo and Longobardi [11]] . If $X$ be a random variable denoting the lifetime of a unit is down at age $t$, then $\bar{X}_t=t-X\mid X<t$ denotes the idle time or inactivity time or reversed residual life of the unit at age $t$.\\
In case of forensic science, people may be interested in estimating $\bar{X}_t$ in order to ascertain the exact time of death of a person. In Insurance industry, it represents the period remained unpaid by a policy holder due to death. For details, see Block, Savits, and Singh[4], Chandra and Roy[7], Maiti and Nanda[18], and Nanda, Singh, Misra, and Paul[25].
The $r^{th}$ moment of $\bar{X}_t$ (for $r=1,~2,~...$) is given by
\begin{eqnarray*}
	\bar{m}_r(t)=E[\bar{X}_t]&=& \frac{1}{F(t)}\int_0^{t}(t-x)^rdF(x)\\&=&\frac{1}{F(t)}\int_0^{t}(t-x)^rf(x,\Phi)dx.
\end{eqnarray*}
In particular, if $r=1$, then $\bar{m}_1(t)$ represents a function called the mean idle time or inactivity time (MIT) or reversed residual life (MRRL) function that indicates the mean inactive life length for a unit which is first observed down at age $t$. The properties of MIT function have been explored by Ahmad, Kayid, and Pellerey[1] and Kayid and Ahmad[16].\\

{\bf Example 3.8:-}\\
{\it Consider the Odds Lindley - Exponential distribution discussed in subsection 2.2.} 
\begin{eqnarray*}
\bar{m}_r(t)&=&\frac{e^\lambda}{1+ \lambda- (1+ \lambda e^{\theta x})e^{-\lambda(e^{\theta x}-1)}}\sum_{j=0}^r \frac{(-1)^j}{\theta ^j}{{r}\choose j}t^{r-j}\sum_{k=0}^j(-1)^{j-k}\binom{j}{k}\left(\ln{\lambda}\right)^{j-k}\nonumber\\&&.\left[\gamma^{(k)}(2, \lambda e^{\theta t})-\gamma^{(k)}(2, \lambda)\right]
\end{eqnarray*}
For the MRRL function,
\begin{eqnarray*}
\bar{m}_1(t)&=&\frac{e^\lambda}{1+ \lambda- (1+ \lambda e^{\theta t})e^{-\lambda(e^{\theta t}-1)}}\left[(t+\frac{\ln\lambda}{\theta})\{\gamma(2,\lambda)-\gamma(2,\lambda e^{\theta t})\}-\frac{1}{\theta}\{\gamma^{(1)}(2,\lambda)-\gamma^{(1)}(2,\lambda e^{\theta t})\}\right].
\end{eqnarray*}

{\bf Example 3.9:-}\\
{\it Consider the Odds Lindley - Pareto distribution discussed in subsection 2.3.} 
\begin{eqnarray*}
\bar{m}_r(t)&=&\frac{e^\lambda}{1+\lambda- [1+ \lambda (\frac{t}{a})^\theta]e^{-\lambda((\frac{t}{a})^\theta-1)}}\sum_{j=0}^r (-1)^j{{r}\choose j}t^{r-j}\frac{a^j}{\lambda^\frac{j}{\theta}}\left[\Gamma(\frac{j}{\theta}+2,\lambda)-\Gamma(\frac{j}{\theta}+2,\frac{\lambda t^\theta}{a^\theta})\right].
\end{eqnarray*}
For the MRRL function,
\begin{eqnarray*}
\bar{m}_1(t)&=&\frac{e^\lambda}{1+\lambda- [1+ \lambda (\frac{t}{a})^\theta]e^{-\lambda((\frac{t}{a})^\theta-1)}}\left[t\{\Gamma(2,\lambda)-\Gamma(2,\frac{\lambda t^\theta}{a^\theta})\}-\frac{a}{\lambda^\frac{1}{\theta}}\{\Gamma(2+\frac{1}{\theta},\lambda)-\Gamma(2+\frac{1}{\theta},\frac{\lambda t^\theta}{a^\theta})\}\right].
\end{eqnarray*}

\begin{figure*}
  \includegraphics[width=1\textwidth]{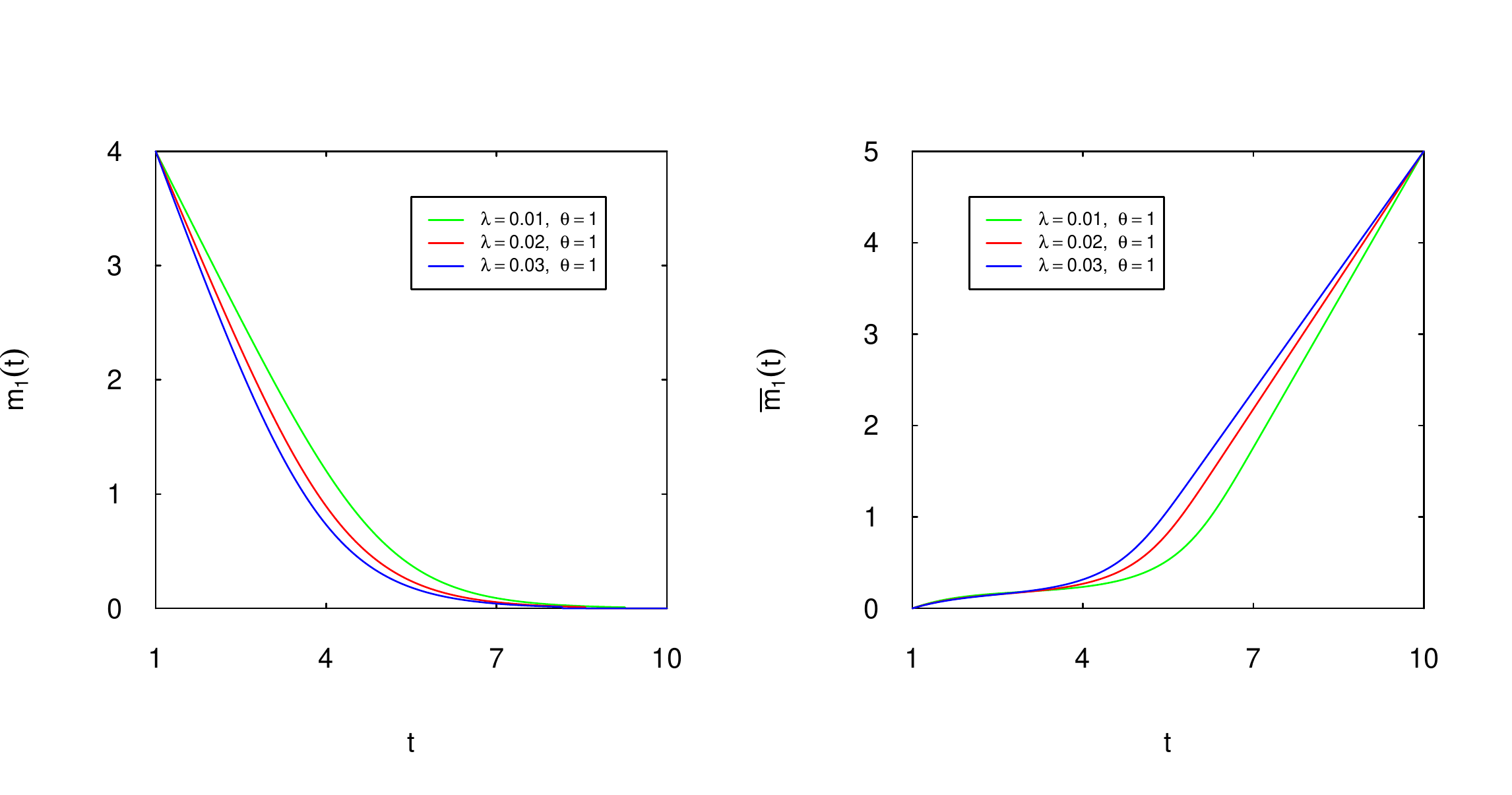}
\caption{Mean Residual Life and Reversed Mean Residual Life  of the Odds Lindley - Exponential distribution}
\label{fig9}
\end{figure*}

\begin{figure*}
  \includegraphics[width=1\textwidth]{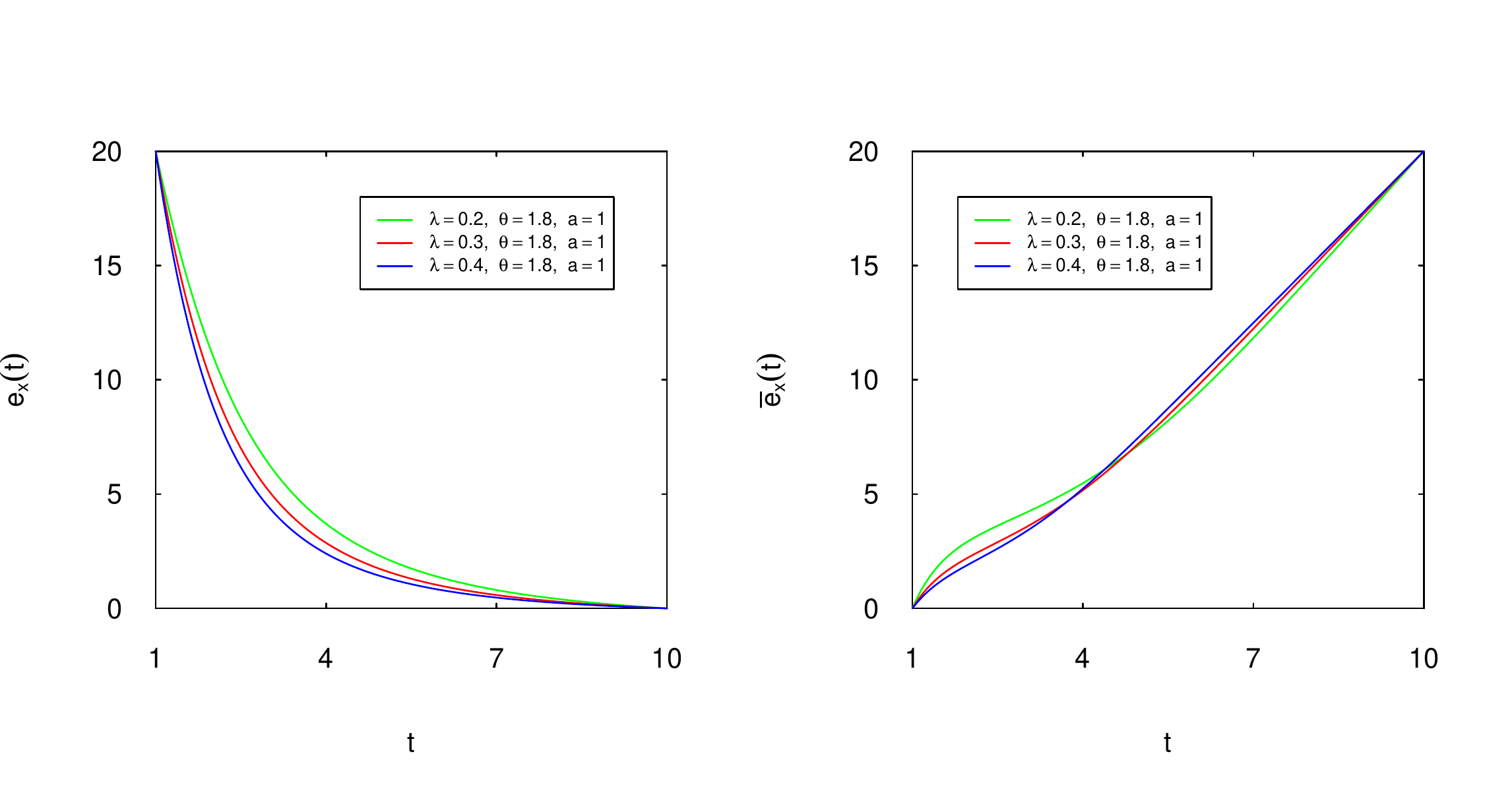}
\caption{Mean Residual Life and Reversed Mean Residual Life  of the Odds Lindley - Pareto distribution}
\label{fig10}
\end{figure*}

\section{Maximum Likelihood Estimation}
In this section, we determine the maximum likelihood estimates(MLEs) of the model parameters of the new family of distributions from complete samples only. Let $ x_1, x_2,....,x_n $ be a observed values from the Odds OPPE -G family of distributions with parameters $\lambda$ and $\xi$. Let $\Phi=(\lambda, \xi)^{T}$ be the p x 1 parameter vector. The log-likelihood function for $\Phi$ is given by

\begin{eqnarray*}
l(\Phi)&=&n\ln h(\lambda) + \sum_{i=0}^{n} \ln \left\lbrace \sum_{k=0}^{s}a_{k}\frac{g(x;\xi)}{[\bar{G}(x;\xi)]^2}V^{k}(x_i;\xi)e^{-\lambda V(x_i;\xi)}\right\rbrace ,
\end{eqnarray*}
where $V(x_i;\xi)=G(x_i;\xi)/\bar{G}(x_i;\xi)$. The components of the score function $U\left(\Phi\right)=\left(U_{\lambda}, U_{\xi}\right)^{T}$ are
\begin{eqnarray*}
U_{\lambda}=\frac{n\frac{\partial}{\partial \lambda}h(\lambda)}{h(\lambda)}+\frac{\partial}{\partial \lambda}\sum_{i=0}^{n} \ln \left\lbrace \sum_{k=0}^{s}a_{k}\frac{g(x;\xi)}{[\bar{G}(x;\xi)]^2}V^{k}(x_i;\xi)e^{-\lambda V(x_i;\xi)}\right\rbrace
\end{eqnarray*}
and
\begin{eqnarray*}
U_{\xi}&=&\frac{\partial}{\partial \xi}\sum_{i=0}^{n} \ln \left\lbrace \sum_{k=0}^{s}a_{k}\frac{g(x;\xi)}{[\bar{G}(x;\xi)]^2}V^{k}(x_i;\xi)e^{-\lambda V(x_i;\xi)}\right\rbrace.
\end{eqnarray*}
Setting $U_{\lambda}$ and $U_{\xi}$ equal to zero and solving the equations simultaneously yields the MLE $\hat{\Phi}=\left(\hat{\lambda},\hat{\xi}\right)^{T}$ of $\Phi=\left(\lambda,\xi\right)^{T}$. These equations cannot be solved analytically and statistical software can be used to solve them numerically using iteration methods such as the Newton- Raphson type algorithms.

\section{Simulation Study}
The direct application of Monte Carlo Simulation Technique for generating random data from the Odds OPPE - G family of distribution fails because the equation F(x) = u, where u is an observation from the uniform distribution on (0, 1), cannot be explicitly solved in x. 

To generate random samples $X_i$, i = 1, 2, 3, .... n, we can use the following algorithm:

\begin{enumerate}
\item Generate $U_i$ $\sim$ Uniform(0, 1), i = 1(1)n
\item If $\frac{\sum_{k=0}^{j-1}a_{k}\frac{k!}{\lambda^{k+1}}}{\sum_{k=0}^{s}a_{k}\frac{k!}{\lambda^{k+1}}} < U_{i}\leq \frac{\sum_{k=0}^{j}a_{k}\frac{k!}{\lambda^{k+1}}}{\sum_{k=0}^{s}a_{k}\frac{k!}{\lambda^{k+1}}}$, i = 1(1)s, then set $Z_i$ = $W_i$, where $W_i$ $\sim$ gamma(j+1, $\lambda$).
\item If $U_i\leq \frac{a_{0}\frac{1}{\lambda}}{\sum_{k=0}^{s}a_{k}\frac{k!}{\lambda^{k+1}}}$, then set $Z_i$ = $V_i$, where $V_i$ $\sim$ exponential($\lambda$).
\end{enumerate}
After using the odds functional form of $G(x;\xi)$, we get the ultimate random data.
For Odds OPPE - Uniform model, set $X_i=\theta Z_i/(1+Z_i)$. For Odds OPPE - Exponential model, set $X_i=\log(1+Z_i)/\theta$. For Odds OPPE - Pareto model, set $X_i=a(1+Z_i)^{\frac{1}{\theta}}$, and for Odds OPPE - Burr XII model, set $X_i=[(1+Z_i)^{\frac{1}{\theta}}-1]^{\frac{1}{\alpha}}$.

Here we assume $s=1$, $a_0=1$, $a_1=1$ to get odds Lindley- Uniform,  odds Lindley- Exponential, odds Lindley- Pareto, and odds Lindley- Burr XII distribution.

A Monte Carlo simulation study was carried out 1000 (=N) times for selected values of n, $\lambda$, $\alpha$, and $\theta$.
\\
(a) Simulation study for Odds Lindley - Uniform distribution, for first simulation, samples of sizes 20, 40, and 100 were considered and values of $\lambda$ were taken as 0.5, 1, 1.5, 3, and 6 for fixed $\theta$=0.1. For second simulation, samples of sizes 20, 40, and 100 were considered and values of $\theta$ were taken as 0.1, 0.5, 1.0, 1.5, and 3 for fixed $\lambda$=0.1. \\
(b) Simulation study for Odds Lindley - Exponential distribution, for first simulation, samples of sizes 20, 40, and 100 were considered and values of $\lambda$ were taken as 0.1, 0.5, 1.5, 3, and 6 for fixed $\theta$=0.1. For second simulation, samples of sizes 20, 40, and 100 were considered and values of $\theta$ were taken as 0.01, 0.5, 1.0, 1.5, and 3 for fixed $\lambda$=0.1. \\
(c) Simulation study for Odds Lindley - Pareto distribution, samples of sizes 20, 40, and 100 were considered and different values of $\lambda$, $\theta$ and $a$ were considered.\\
(d) Simulation study for Odds Lindley - Burr XII distribution, samples of sizes 20, 40, and 100 were considered and different values of $\lambda$, $\theta$ and $\alpha$ were considered.\\
The required numerical evaluations are carried out using R 3.1.1 software. The following two measures were computed:

\begin{enumerate}
\item Bias of the simulated estimates $\hat{\lambda}$, $\hat{\alpha}$ and $\hat{\theta}$, for i=1, 2, 3, .....,N:\\
$\frac{1}{N}\sum_{i=1}^{N}(\hat{\lambda_{i}}-\lambda)$, $\frac{1}{N}\sum_{i=1}^{N}(\hat{\alpha_{i}}-\alpha)$ and $\frac{1}{N}\sum_{i=1}^{N}(\hat{\theta_{i}}-\theta)$,
\item Mean Square Error (MSE) of the simulated estimates $\hat{\lambda}$, $\hat{\alpha}$ and $\hat{\theta}$, for i=1, 2, 3, .....,N:\\
$\frac{1}{N}\sum_{i=1}^{N}(\hat{\lambda_{i}}-\lambda)^{2}$, $\frac{1}{N}\sum_{i=1}^{N}(\hat{\alpha_{i}}-\alpha)^2$ and $\frac{1}{N}\sum_{i=1}^{N}(\hat{\theta_{i}}-\theta)^{2}$.
\end{enumerate}

The result of the simulation study for Odds Lindley - Uniform distribution has been tabulated in Table 2. It shows that\\
(i) Bias and MSE decreases as n increases. \\
(ii) Bias and MSE increases as the values of $\lambda$ increases for fixed $\theta$=0.1. \\
(iii) Bias and MSE increases as the values of $\theta$ increases for fixed $\lambda$=0.1.\\

The result of the simulation study for Odds Lindley - Exponential distribution has been tabulated in Table 3. It shows that\\
(i) Bias and MSE decreases as n increases. \\
(ii) Bias and MSE increases as the values of $\lambda$ increases for fixed $\theta$=0.1. \\
(iii) Bias and MSE increases as the values of $\theta$ increases for fixed $\lambda$=0.1.\\

The result of the simulation study for Odds Lindley - Pareto distribution has been tabulated in Table 4. It shows that\\
(i) Bias and MSE decreases as n increases. \\
(ii) Bias and MSE increases as the values of $\lambda$ and $\theta$ increases for fixed $a$=0.1. \\
(iii) Bias and MSE increases as the values of $\lambda$ and $a$ increases for fixed $\theta$=1.

The result of the simulation study for Odds Lindley - Burr XII distribution has been tabulated in Table 5. It shows that\\
(i) Bias and MSE decreases as n increases. \\
(ii) Bias and MSE increases as the values of $\theta$ and $\alpha$ increases for fixed $\lambda$=0.1. \\
(iii) Bias and MSE increases as the values of $\lambda$ and $\alpha$ increases for fixed $\theta$=0.1.

\begin{table}
\caption{Average Bias and MSE of the estimator of $\hat{\lambda}$ and $\hat{\theta}$ for Odds Lindley - Uniform distribution}
\label{tab:2}  
\begin{tabular}{|c|c|c|c|c|}\hline
&\multicolumn{2}{|c|}{$\hat{\lambda}=0.5$}& \multicolumn{2}{|c|}{$\hat{\theta}=0.1$}\\\hline
n &	Bias &	MSE	& Bias & MSE\\ \hline
20&	-0.2776&	0.0900&	-0.0097&	0.0001\\ \hline
40&	-0.2754&	0.0852&	-0.0082&	0.0001\\ \hline
100& -0.2646&	0.0766&	-0.0070&	0.0001\\ \hline
&\multicolumn{2}{|c|} {$\hat{\lambda}=1$}&\multicolumn{2}{|c|}{$\hat{\theta}=0.1$}\\ \hline
n&	Bias &	MSE	&Bias&	MSE\\ \hline
20&	-0.5729&	0.3665&	-0.0184&	0.0004\\ \hline
40&	-0.5638&	0.3661&	-0.0160&	0.0003\\ \hline
100&	-0.5513&	0.3301&	-0.0136&	0.0002\\ \hline
&\multicolumn{2}{|c|}{$\hat{\lambda}=1.5$}&\multicolumn{2}{|c|}{$\hat{\theta}=0.1$}\\ \hline
n&	Bias &	MSE	&Bias&	MSE\\ \hline
20&	-0.8772&	0.8744&	-0.0260&	0.0007\\ \hline
40&	-0.8749&	0.8414&	-0.0225&	0.0005\\ \hline
100&	-0.8422&	0.7634&	-0.0195&	0.0004\\ \hline
&\multicolumn{2}{|c|}{$\hat{\lambda}=3$}&\multicolumn{2}{|c|}{$\hat{\theta}=0.1$}\\ \hline
n&	Bias &	MSE	&Bias&	MSE\\ \hline
20&	-1.9486&	4.1142&	-0.0423&	0.0018\\ \hline
40&	-1.9478&	4.0112&	-0.0383&	0.0015\\ \hline
100&	-1.8676&	3.6703&	-0.0338&	0.0012\\ \hline
&\multicolumn{2}{|c|}{$\hat{\lambda}=6$}&\multicolumn{2}{|c|}{$\hat{\theta}=0.1$}\\ \hline
n&	Bias &	MSE	&Bias&	MSE\\ \hline
20&	-4.4572&	20.688&	-0.0608&	0.0038\\ \hline
40&	-4.4059&	19.969&	-0.0562&	0.0032\\ \hline
100&	-4.2437&	18.447&	-0.0516&	0.0027\\ \hline
\end{tabular}
\quad
\begin{tabular}{|c|c|c|c|c|}
\hline
&\multicolumn{2}{|c|}{$\hat{\lambda}=0.1$}&\multicolumn{2}{|c|}{$\hat{\theta}=0.1$}\\\hline
n&	Bias &	MSE	&Bias&	MSE \\ \hline
20&	-0.0561&	0.0036&	-0.0020&	0.0000\\ \hline
40&	-0.0557&	0.0035&	-0.0017&	0.0000\\ \hline
100&	-0.0536&	0.0031&	-0.0014&	0.0000\\ \hline
&\multicolumn{2}{|c|}{$\hat{\lambda}=0.1$}&\multicolumn{2}{|c|}{$\hat{\theta}=0.5$}\\ \hline
n&	Bias &	MSE	&Bias&	MSE\\ \hline
20&	-0.0568&	0.0037&	-0.0097&	0.0001\\ \hline
40&	-0.0565&	0.0036&	-0.0085&	0.0001\\ \hline
100&	-0.0533&	0.0031&	-0.0071&	0.0001\\ \hline
&\multicolumn{2}{|c|}{$\hat{\lambda}=0.1$}&\multicolumn{2}{|c|}{$\hat{\theta}=1$}\\ \hline
n&	Bias &	MSE	&Bias&	MSE\\ \hline
20&	-0.0556&	0.0036&	-0.0195&	0.0004\\ \hline
40&	-0.0555&	0.0035&	-0.0166&	0.0003\\ \hline
100&	-0.0529&	0.0031&	-0.0142&	0.0002\\ \hline
&\multicolumn{2}{|c|}{$\hat{\lambda}=0.1$}&\multicolumn{2}{|c|}{$\hat{\theta}=1.5$}\\ \hline
n&	Bias &	MSE	&Bias&	MSE\\ \hline
20&	-0.0576&	0.0038&	-0.0298&	0.0009\\ \hline
40&	-0.0551&	0.0034&	-0.0253&	0.0007\\ \hline
100&	-0.0526&	0.0030&	-0.0213&	0.0005\\ \hline
&\multicolumn{2}{|c|}{$\hat{\lambda}=0.1$}&\multicolumn{2}{|c|}{$\hat{\theta}=3$}\\ \hline
n&	Bias &	MSE	&Bias&	MSE\\ \hline
20&	-0.0563&	0.0037&	-0.0591&	0.0037\\ \hline
40&	-0.0555&	0.0034&	-0.0503&	0.0027\\ \hline
100&	-0.0529&	0.0031&	-0.0426&	0.0019\\ \hline
\end{tabular}
\end{table}

\begin{table}
\caption{Average Bias and MSE of the estimator of $\hat{\lambda}$ and $\hat{\theta}$ for Odds Lindley - Exponential distribution}
\label{tab:3}  
\begin{tabular}{|c|c|c|c|c|}\hline
&\multicolumn{2}{|c|}{$\hat{\lambda}=0.1$}& \multicolumn{2}{|c|}{$\hat{\theta}=0.1$}\\\hline
n &	Bias &	MSE	& Bias & MSE\\ \hline
20&	0.0160&	0.0026&	0.0297&	0.0012\\ \hline
40&	0.0142&	0.0017&	0.0269&	0.0008\\ \hline
100& 0.0125&	0.0008&	0.0250&	0.0007\\ \hline
&\multicolumn{2}{|c|} {$\hat{\lambda}=0.5$}&\multicolumn{2}{|c|}{$\hat{\theta}=0.1$}\\ \hline
n&	Bias &	MSE	&Bias&	MSE\\ \hline
20&	-0.1501&	0.0361&	0.0715&	0.0057\\ \hline
40&	-0.1414&	0.0270&	0.0667&	0.0047\\ \hline
100&	-0.1321&	0.0203&	0.0631&	0.0041\\ \hline
&\multicolumn{2}{|c|}{$\hat{\lambda}=1.5$}&\multicolumn{2}{|c|}{$\hat{\theta}=0.1$}\\ \hline
n&	Bias &	MSE	&Bias&	MSE\\ \hline
20&	-0.8637&	0.7744&	0.1610&	0.0286\\ \hline
40&	-0.8432&	0.7272&	0.1489&	0.0234\\ \hline
100&	-0.8184&	0.6767&	0.1395&	0.0199\\ \hline
&\multicolumn{2}{|c|}{$\hat{\lambda}=3$}&\multicolumn{2}{|c|}{$\hat{\theta}=0.1$}\\ \hline
n&	Bias &	MSE	&Bias&	MSE\\ \hline
20&	-2.2043&	4.8961&	0.2995&	0.0991\\ \hline
40&	-2.1562&	4.6724&	0.2688&	0.0762\\ \hline
100&	-2.1208&	4.5102&	0.2529&	0.0657\\ \hline
&\multicolumn{2}{|c|}{$\hat{\lambda}=6$}&\multicolumn{2}{|c|}{$\hat{\theta}=0.1$}\\ \hline
n&	Bias &	MSE	&Bias&	MSE\\ \hline
20&	-5.2467&	27.879&	0.6196&	0.4522\\ \hline
40&	-5.0995&	26.209&	0.5408&	0.3258\\ \hline
100&	-4.9878&	24.955&	0.4782&	0.2396\\ \hline
\end{tabular}
\quad
\begin{tabular}{|c|c|c|c|c|}
\hline
&\multicolumn{2}{|c|}{$\hat{\lambda}=0.1$}&\multicolumn{2}{|c|}{$\hat{\theta}=0.01$}\\\hline
n&	Bias &	MSE	&Bias&	MSE \\ \hline
20&	0.0137&	0.0027&	0.0031&	0.0000\\ \hline
40&	0.0111&	0.0014&	0.0028&	0.0000\\ \hline
100&	0.0104&	0.0007&	0.0026&	0.0000\\ \hline

&\multicolumn{2}{|c|}{$\hat{\lambda}=0.1$}&\multicolumn{2}{|c|}{$\hat{\theta}=0.5$}\\ \hline
n&	Bias &	MSE	&Bias&	MSE\\ \hline
20&	0.0141&	0.0050&	0.1576&	0.0326\\ \hline
40&	0.0116&	0.0014&	0.1388&	0.0223\\ \hline
100&	0.0053&	0.0007&	0.1287&	0.0177\\ \hline

&\multicolumn{2}{|c|}{$\hat{\lambda}=0.1$}&\multicolumn{2}{|c|}{$\hat{\theta}=1$}\\ \hline
n&	Bias &	MSE	&Bias&	MSE\\ \hline
20&	0.0158&	0.0069&	0.3091&	0.1306\\ \hline
40&	0.0114&	0.0026&	0.2755&	0.0880\\ \hline
100&	0.0070&	0.0008&	0.2510&	0.0673\\ \hline

&\multicolumn{2}{|c|}{$\hat{\lambda}=0.1$}&\multicolumn{2}{|c|}{$\hat{\theta}=1.5$}\\ \hline
n&	Bias &	MSE	&Bias&	MSE\\ \hline
20&	0.0147&	0.0052&	0.4628&	0.2815\\ \hline
40&	0.0112&	0.0015&	0.4147&	0.1981\\ \hline
100&	0.0080&	0.0008&	0.3819&	0.1561\\ \hline

&\multicolumn{2}{|c|}{$\hat{\lambda}=0.1$}&\multicolumn{2}{|c|}{$\hat{\theta}=3$}\\ \hline
n&	Bias &	MSE	&Bias&	MSE\\ \hline
20&	-0.3464&	0.3641&	1.0259&	2.2555\\ \hline
40&	-0.3392&	0.3578&	0.9580&	1.9045\\ \hline
100&	-0.3315&	0.3486&	0.8696&	1.6765\\ \hline
\end{tabular}
\end{table}

\begin{table}
\caption{Average Bias and MSE of the estimator of $\hat{\lambda}$, $\hat{\theta}$ and $a$ for Odds Lindley - Pareto distribution}
\begin{center}
\label{tab:4}  
\begin{tabular}{|c|c|c|c|c|c|c|}\hline
&\multicolumn{2}{|c|}{$\hat{\lambda}=1$}& \multicolumn{2}{|c|}{$\hat{\theta}=1$}& \multicolumn{2}{|c|}{$\hat{a}=0.1$}\\\hline
n &	Bias &	MSE	& Bias & MSE & Bias & MSE\\ \hline
20&	0.2650&	0.6347&	0.1656&	0.1885&	0.0092&	0.0002\\ \hline
40&	0.0660&	0.2070&	0.0673&	0.0701&	0.0048&	0.0000\\ \hline
100&	0.0370&	0.0671&	0.0175&	0.0228&	0.0020&	0.0000\\ \hline
&\multicolumn{2}{|c|}{$\hat{\lambda}=0.1$}& \multicolumn{2}{|c|}{$\hat{\theta}=1$}& \multicolumn{2}{|c|}{$\hat{a}=0.1$}\\ \hline
n&	Bias &	MSE	&Bias&	MSE & Bias & MSE\\ \hline
20&	0.2336&	0.1435&	0.0833&	0.0708&	0.2371&	0.0869\\ \hline
40&	0.1213&	0.0622&	0.0432&	0.0307&	0.1440&	0.0322\\ \hline
100&	0.0573&	0.0320&	0.0147&	0.0116&	0.0733&	0.0087\\ \hline
&\multicolumn{2}{|c|}{$\hat{\lambda}=0.5$}& \multicolumn{2}{|c|}{$\hat{\theta}=2$}& \multicolumn{2}{|c|}{$\hat{a}=0.1$}\\ \hline
n&	Bias &	MSE	&Bias&	MSE & Bias & MSE\\ \hline
20&	0.1601&	0.2125&	0.1923&	0.4370&	0.0121&	0.0003\\ \hline
40&	0.0948&	0.1096&	0.0741&	0.1847&	0.0067&	0.0001\\ \hline
100&	0.0346&	0.0365&	0.0382&	0.0619&	0.0028&	0.0000\\ \hline
&\multicolumn{2}{|c|}{$\hat{\lambda}=0.5$}& \multicolumn{2}{|c|}{$\hat{\theta}=2$}& \multicolumn{2}{|c|}{$\hat{a}=0.5$}\\ \hline
n&	Bias &	MSE	&Bias&	MSE & Bias & MSE\\ \hline
20&	0.1396&	0.2372&	0.2277&	0.4636&	0.0589&	0.0060\\ \hline
40&	0.0938&	0.1213&	0.0898&	0.1728&	0.0338&	0.0020\\ \hline
100&	0.0372&	0.0376&	0.0327&	0.0599&	0.0147&	0.0004\\ \hline
&\multicolumn{2}{|c|}{$\hat{\lambda}=1$}& \multicolumn{2}{|c|}{$\hat{\theta}=1$}& \multicolumn{2}{|c|}{$\hat{a}=0.5$}\\ \hline
n&	Bias &	MSE	&Bias&	MSE & Bias & MSE\\ \hline
20&	0.4334&	0.6902&	0.1434&	0.1785&	0.0469&	0.0045\\ \hline
40&	0.0775&	0.2066&	0.0558&	0.0648&	0.0244&	0.0011\\ \hline
100&	0.0311&	0.0808&	0.0268&	0.0270&	0.0098&	0.0002\\ \hline
\end{tabular}
\end{center}
\end{table}

\begin{table}
\caption{Average Bias and MSE of the estimator of $\hat{\lambda}$, $\hat{\theta}$ and $\hat{\alpha}$ for Odds Lindley - Burr XII distribution}
\begin{center}
\label{tab:5}  
\begin{tabular}{|c|c|c|c|c|c|c|}\hline
&\multicolumn{2}{|c|}{$\hat{\lambda}=0.1$}& \multicolumn{2}{|c|}{$\hat{\theta}=0.1$}& \multicolumn{2}{|c|}{$\hat{\alpha}=0.1$}\\\hline
n &	Bias &	MSE	& Bias & MSE & Bias & MSE\\ \hline
20&	-0.0030&	0.0037&	0.0129&	0.0144&	0.0222&	0.0068\\ \hline
40&	-0.0011&	0.0015&	0.0078&	0.0072&	0.0165&	0.0034\\ \hline
100&	-0.0002&	0.0007&	0.0046&	0.0068&	0.0117&	0.0019\\ \hline
&\multicolumn{2}{|c|}{$\hat{\lambda}=0.1$}& \multicolumn{2}{|c|}{$\hat{\theta}=0.5$}& \multicolumn{2}{|c|}{$\hat{\alpha}=0.1$}\\ \hline
n&	Bias &	MSE	&Bias&	MSE & Bias & MSE\\ \hline
20&	-0.0127&	0.0051&	0.1157&	0.5855&	0.1019&	0.0285\\ \hline
40&	-0.0103&	0.0024&	0.1054&	0.4946&	0.0914&	0.0222\\ \hline
100&	-0.0068&	0.0010&	0.0390&	0.2629&	0.0747&	0.0175\\ \hline
&\multicolumn{2}{|c|}{$\hat{\lambda}=0.1$}& \multicolumn{2}{|c|}{$\hat{\theta}=0.5$}& \multicolumn{2}{|c|}{$\hat{\alpha}=0.5$}\\ \hline
n&	Bias &	MSE	&Bias&	MSE & Bias & MSE\\ \hline
20&	-0.0133&	0.0043&	0.1516&	0.7191&	0.6022&	0.8379\\ \hline
40&	-0.0117&	0.0024&	0.0996&	0.5498&	0.5712&	0.7528\\ \hline
100&	-0.0081&	0.0011&	0.0494&	0.3546&	0.4923&	0.6174\\ \hline
&\multicolumn{2}{|c|}{$\hat{\lambda}=0.5$}& \multicolumn{2}{|c|}{$\hat{\theta}=0.1$}& \multicolumn{2}{|c|}{$\hat{\alpha}=0.1$}\\ \hline
n&	Bias &	MSE	&Bias&	MSE & Bias & MSE\\ \hline
20&	-0.0144&	0.0667&	-0.0124&	0.0361&	0.2779&	0.2553\\ \hline
40&	-0.0061&	0.0313&	-0.0087&	0.0135&	0.2356&	0.1801\\ \hline
100&	-0.0060&	0.0109&	-0.0040&	0.0055&	0.1485&	0.0997\\ \hline
&\multicolumn{2}{|c|}{$\hat{\lambda}=0.5$}& \multicolumn{2}{|c|}{$\hat{\theta}=0.1$}& \multicolumn{2}{|c|}{$\hat{\alpha}=0.5$}\\ \hline
n&	Bias &	MSE	&Bias&	MSE & Bias & MSE\\ \hline
20&	-0.0082&	0.0632&	-0.0155&	0.0501&	1.1588&	3.7477\\ \hline
40&	-0.0016&	0.0296&	-0.0118&	0.0178&	1.0481&	2.7680\\ \hline
100&	-0.0010&	0.0125&	-0.0016&	0.0043&	0.6658&	1.5149\\ \hline
\end{tabular}
\end{center}
\end{table}

\section{Application}
\begin{figure*}
\includegraphics[width=.99\textwidth]{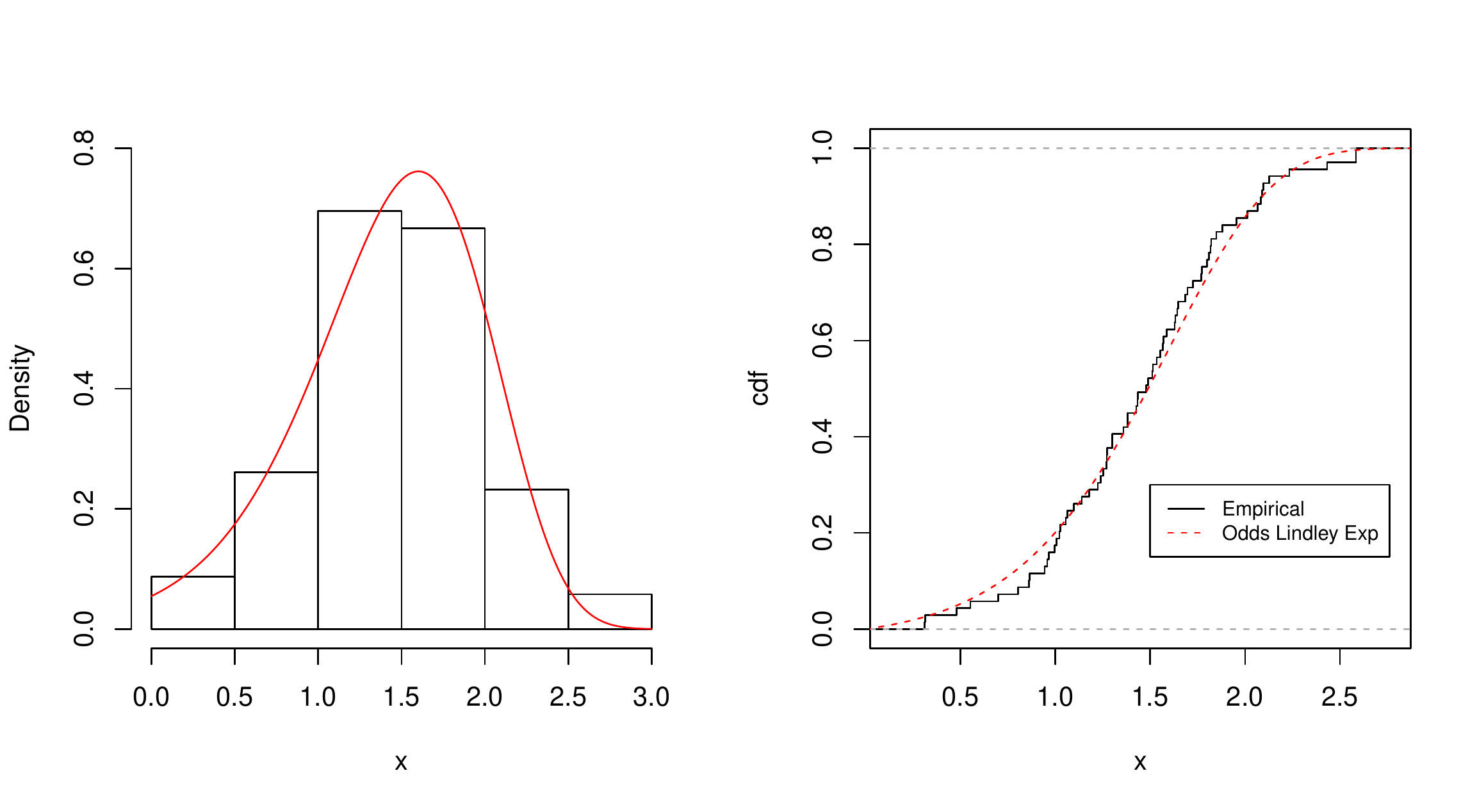} 
\caption{Plots of the estimated pdf and cdf of the Odds Lindley Exponential model for Single Carbon Fibers tested at gauge lengths of 20 mm}
\label{fig11}
\end{figure*}

In this section, we fit the above model to three real data sets.\\
\textbf{Data Set 1:-} The first data set represents the tensile strength data measured in GPa for single-carbon fibers that were tested at gauge lengths of 20 mm. Alzaatreh and Knight (2013) fitted this data to the Gamma-Half Normal distribution. We have fitted this data set with the Odds Lindley Exponential Distribution. The estimated values of the parameters were $\hat{\lambda}=0.2202$ and $\hat{\theta}=1.3773$ and AIC = $104.3232$. In the Odds Lindley Exponential Distribution fitting only two parameters are to be estimated that will minimize estimation error with compared to the Gamma-Half Normal distribution. Histogram and fitted Odds Lindley-Exponential curve to data have been shown in Figure $\ref{fig11}$.

\begin{table}
\caption{Single Carbon Fibers at 20 mm}
\begin{center}
\label{tab:6} 
\begin{tabular}{llllllllllllll}
\hline\noalign{\smallskip}
0.312 0.700 0.944 1.006 1.063 1.224 1.272 1.359 1.434 1.511 1.566 1.633 1.697 1.800 \\
1.848 2.067 2.128 2.585 0.314 0.803 0.958 1.021 1.098 1.240 1.274 1.382 1.435 1.514 \\
1.570 1.642 1.726 1.809 1.880 2.084 2.233 0.479 0.861 0.966 1.027 1.140 1.253 1.301 \\
1.382 1.478 1.535 1.586 1.648 1.770 1.818 1.954 2.090 2.433 0.552 0.865 0.997 1.055 \\
1.179 1.270 1.301 1.426 1.490 1.554 1.629 1.684 1.773 1.821 2.012 2.096 2.585\\
\noalign{\smallskip}\hline
\end{tabular}
\end{center}
\end{table}

\begin{table}
\caption{Summarized results of fitting different distributions for Single Carbon Fibers tested at gauge lengths of 20 mm}
\begin{center}
\label{tab:7} 
\begin{tabular}{lll}
\hline\noalign{\smallskip}
Distribution & Estimate of the parameters & AIC  \\
\noalign{\smallskip}\hline\noalign{\smallskip}
Gamma-Half Normal Distribution & $\hat{\theta}=0.3934, \hat{\alpha}=2.8794, \hat{\beta}=3.1725$ & $105.3572$\\
Odds Lindley-Exponential Distribution & $\hat{\lambda}=0.2202, \hat{\theta}=1.3773$ & $104.3232$\\
\noalign{\smallskip}\hline
\end{tabular}
\end{center}
\end{table}

\begin{figure*}
\includegraphics[width=.99\textwidth]{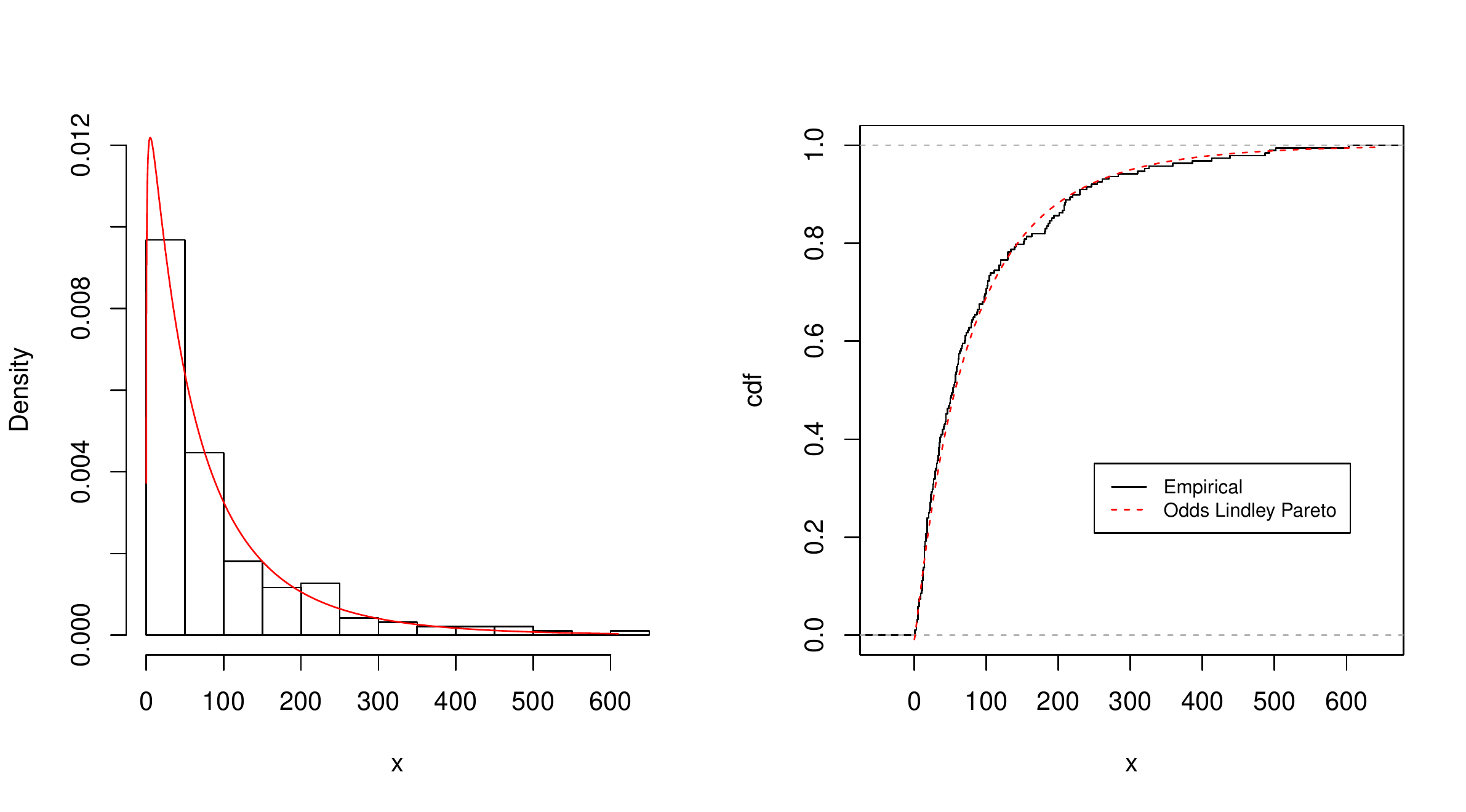} 
\caption{Plots of the estimated pdf and cdf of the Odds Lindley-Pareto model for the number of successive failures for the air conditioning system}
\label{fig12}
\end{figure*}

\textbf{Data Set 2:-}
The second data set is the number of failures for the air conditioning system of jet airplanes. These data were reported by Cordeiro and Lemonte (2011) and Huang and Oluyede (2014):194, 413, 90, 74, 55, 23, 97, 50, 359, 50, 130, 487, 57, 102, 15, 14, 10, 57, 320, 261, 51, 44, 9, 254, 493, 33, 18, 209, 41, 58, 60, 48, 56, 87, 11, 102, 12, 5, 14, 14, 29, 37, 186, 29, 104, 7, 4, 72, 270, 283, 7, 61, 100, 61, 502, 220, 120, 141, 22, 603, 35, 98, 54, 100, 11, 181, 65, 49, 12, 239, 14, 18, 39, 3, 12, 5, 32, 9, 438, 43, 134, 184, 20, 386, 182, 71, 80, 188, 230, 152, 5, 36, 79, 59, 33, 246, 1, 79, 3, 27, 201, 84, 27, 156, 21, 16, 88, 130, 14, 118, 44, 15, 42, 106, 46, 230, 26, 59, 153, 104, 20, 206, 5, 66, 34, 29, 26, 35, 5, 82, 31, 118, 326, 12, 54, 36, 34, 18, 25, 120, 31, 22, 18, 216, 139, 67, 310, 3, 46, 210, 57, 76, 14, 111, 97, 62, 39, 30, 7, 44, 11, 63, 23, 22, 23, 14, 18, 13, 34, 16, 18, 130, 90, 163, 208, 1, 24, 70, 16, 101, 52, 208, 95, 62, 11, 191, 14, 71. Some descriptive statistics for these data are given below. Histogram shows that the data set is positively skewed. Thiago A. N. de Andrade, Marcelo Bourguignon, Gauss M. Cordeiro (2016) fitted this data to the exponentiated generalized extended exponential distribution(EGEE). We have fitted this data set with the Odds Lindley-Pareto distribution. The estimated values of the parameters were $\lambda= 0.1395$, $\theta= 0.6183$, $a= 1$, log-likelihood =$ -1023.159$ and AIC = $2052.319$. Histogram and fitted Odds Lindley Pareto curve to data have been shown in Figure $\ref{fig12}$.\\

\begin{table}
\caption{Summarized results of fitting different distributions to data set of Cordeiro and Lemonte (2011) and Huang and Oluyede (2014)}
\begin{center}
\label{tab:8} 
\begin{tabular}{llll}
\hline\noalign{\smallskip}
Distribution & Estimate of the parameters & AIC  \\
\noalign{\smallskip}\hline\noalign{\smallskip}
EGEE Distribution & $\hat{a}=0.0639, \hat{b}=0.647, \hat{\alpha}=0.1497, \hat{\beta}=183.90$ & $2077.400$\\
Odds Lindley-Pareto Distribution & $\hat{\lambda}=0.1395, \hat{\theta}=0.6183, \hat{a}=1$ & $2052.319$\\
\noalign{\smallskip}\hline
\end{tabular}
\end{center}
\end{table}

\begin{figure*}
\includegraphics[width=.99\textwidth]{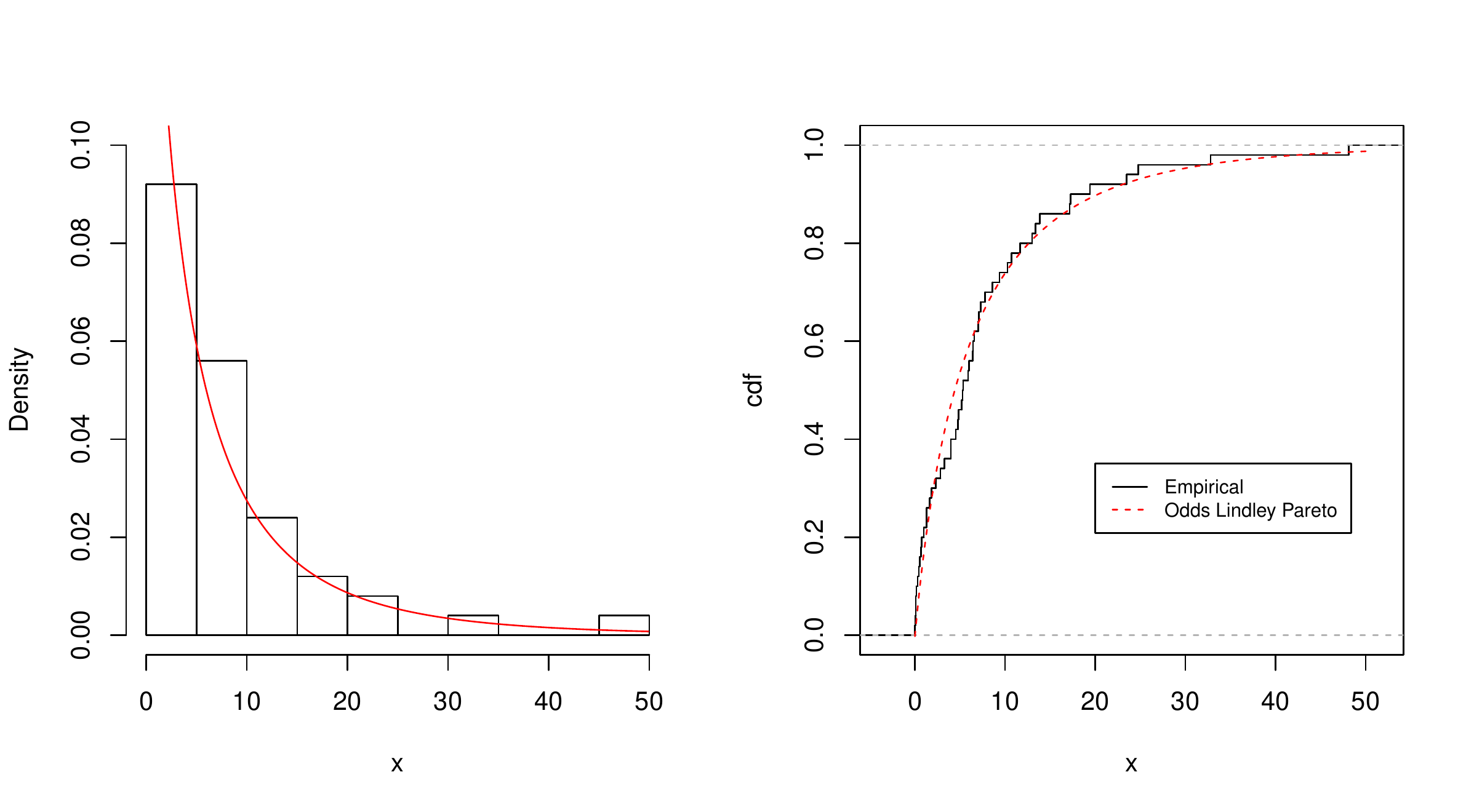} 
\caption{Plots of the estimated pdf and cdf of the Odds Lindley-Pareto model for the failure times of 50 components}
\label{fig13}
\end{figure*}

\textbf{Data Set 3:-} The third data set was represented by Murthy et al. (2004) on the failure times (in weeks) of 50 components. The data are: 0.013, 0.065, 0.111, 0.111, 0.163, 0.309, 0.426, 0.535, 0.684, 0.747, 0.997, 1.284, 1.304, 1.647, 1.829, 2.336, 2.838, 3.269, 3.977, 3.981, 4.520, 4.789, 4.849, 5.202, 5.291, 5.349, 5.911, 6.018, 6.427, 6.456, 6.572, 7.023, 7.087, 7.291, 7.787, 8.596, 9.388, 10.261, 10.713, 11.658, 13.006, 13.388, 13.842, 17.152, 17.283, 19.418, 23.471, 24.777, 32.795, 48.105. Histogram shows that the data set is positively skewed. Thiago A. N. de Andrade, Marcelo Bourguignon, Gauss M. Cordeiro (2016) fitted this data to the exponentiated generalized extended exponential distribution(EGEE). We have fitted this data set with the Odds Generalized Lindley-Pareto distribution. The estimated values of the parameters were $\lambda= 0.0682$, $\theta= 0.5499$, $a= 0.013$, log-likelihood =$-150.196$ and AIC = $306.391$. Histogram and fitted Lindley Pareto curve to data have been shown in Figure $\ref{fig13}$.

\begin{table}
\caption{Summarized results of fitting different distributions to data set of Cordeiro and Lemonte (2011) and Huang and Oluyede (2014)}
\begin{center}
\label{tab:9} 
\begin{tabular}{llll}
\hline\noalign{\smallskip}
Distribution & Estimate of the parameters & AIC  \\
\noalign{\smallskip}\hline\noalign{\smallskip}
EGEE Distribution & $\hat{a}=0.3659, \hat{b}=0.3103, \hat{\alpha}=0.3239, \hat{\beta}=0.6041$ & $308.300$\\
Odds Lindley-Pareto Distribution & $\hat{\lambda}=0.0682, \hat{\theta}=0.5499, \hat{a}=0.013$ & $306.391$\\
\noalign{\smallskip}\hline
\end{tabular}
\end{center}
\end{table}

\section{Concluding Remarks}
We have introduced and studied a new generalized family of distributions, called the Odds OPPE - G Family of distributions. Properties of the Odds OPPE - G Family of distributions include: an expansion for the density function and expressions for the quantile function, moment generating function, ordinary moments, incomplete moments, mean deviations, Lorenz and Benferroni curves, reliability properties including mean residual life and mean inactivity time, and order statistics. The maximum likelihood method is employed to estimate the model parameters. Three real data sets are used to demonstrate the flexibility of distribution belonging to the introduced family. The special models give better fits than other models. It is expected that the findings of the paper will be quite useful for the practitioners in various fields of probability, statistics and applied sciences.
\par The Transmuted OPPE distribution and its properties study and applications are in progress. The estimation aspect of the pdf and cdf of the OPPE distribution is also in pipeline and the progress is to be communicated shortly.

\section*{References}
\begin{enumerate}

\item Ahmad I A, Kayid M, Pellerey F (2005). "Further results involving the MIT order and the IMIT class". \textit{Probability in the Engineering and Informational Sciences}, \textbf{19}, 377-395.
\item Alzaatreh A, Lee C, Famoye F (2013). "A New Method for Generating Families of Distributions". \textit{Metron}, \textbf{71}, 63-79.
\item Azzalini A (1985). "A Class of Distributions which Includes the Normal Ones". \textit{Scand J Stat}., \textbf{12}, 171-178.
\item Bhattacharya R, Maiti S S, Choudhury M M and Mukherjee D (2020). "Minimum Variance Unbiased Estimation of Reliability Function
for a Class OF Generalizations of
Lindley Distribution". \textit{Calcutta Statistical Association Bulletin}, \textbf{In press}.
\item Block H W, Savits T H, Singh H (1998). "The reversed hazard rate function". \textit{Probability in Engineering and Informational Sciences}, \textbf{12}, 69-90.
\item Bourguignon M, Silva R B, Cordeiro G M (2014). "The Weibull-G Family of Probability Distributions". \textit{Journal of Data Science}, \textbf{12}, 53-68.
\item Burr I W (1942). "Cumulative frequency functions". \textit{Ann. Math. Stat}., \textbf{13}, 215-232.
\item Chandra N K, Roy D (2001). "Some results on reversed hazard rate". \textit{Probability in the Engineering and Informational Sciences}, \textbf{15}(1), 95-102.
\item Cordeiro G M, Barreto-Souza (2009). "General Results for a Class of Beta G Distributions". Unpublished material.
\item Cordeiro G M, deCastro M (2011). "A New Family of Generalized Distributions". \textit{Journal of Statistical Computation and Simulation}, \textbf{81}, 883-893.
\item David H A, Nagaraja H N (2003). \textit{Order Statistics}. John Wiley \& Sons, New Jersey.
\item Di Crescenzo A, Longobardi M (2002). "Entropy-based measure of uncertainty in past lifetime distributions". \textit{Journal of Applied Probability}, \textbf{39}, 434-440.
\item Eugene N, Lee C, Famoye F (2002). "Beta-normal Distribution and Its Applications". \textit{Journal of Applied Probability}, \textbf{31}, 497-512.
\item Ghitany M E, Atieh B, Nadarajah S (2008). "Lindley Distribution and Its Applications". \textit{Mathematics and Computers in Simulation}, \textbf{78}, 493-506.
\item Johnson N L (1949). Systems of frequency curves generated by methods of translation". \textit{Biometrika}, \textbf{36}, 149-176.
\item JonesMC (2009). Kumaraswamy's Distribution: "A Beta-Type Distribution with Tractability Advantages". \textit{Statistical Methodology}, \textbf{6}, 70-81.
\item Kayid M, Ahmad I A (2004). "On the mean inactivity time ordering with reliability applications". \textit{Probability in the Engineering and Informational Sciences}, \textbf{18}(03), 395-409.
\item Kotz S, Lai C D, Xie M (2003). "On the Effect of Redundancy for Systems with Dependent Components". \textit{IIE Trans}, \textbf{35}, 1103-1110.
\item Maiti S S, Nanda A K (2009). "A loglikelihood-based shape measure of past lifetime distribution". \textit{Calcutta Statistical Association Bulletin}, \textbf{61}, 303-320.
\item Maiti S S, Pramanik S (2015). "Odds Generalized Exponential-Exponential Distribution". \textit{Journal of Data Science}, \textbf{13}, 733-754.
\item Maiti S S, Pramanik S (2016a). "Odds Generalized Exponential Uniform Distribution and its Application". \textit{Research \& Reviews: Journal of Statistics}, \textbf{5}(1), 33-45.
\item Maiti S S, Pramanik S (2016b). "Odds Generalized Exponential-Pareto Distribution: Properties and Application". \textit{Pakistan Journal of Statistics and Operations Research}, \textbf{12}(2), 257-279.
\item Maiti S S, Pramanik S (2018). "Odds Xgamma – G Family of Distributions". \textit{IAPQR Transactions}, \textbf{43}(2), 135-163.
\item Marshall A N, Olkin I (1997). "A New Method for Adding a Parameter to a Family of Distributions with Applications to the Exponential and Weibull Families". \textit{Biometrika}, \textbf{84}, 641-552.
\item McDonald J B (1984). "Some Generalized Functions for the Size Distribution of Income". \textit{Biometrika}, \textbf{52}, 647-663.
\item Nanda A K, Singh H, Misra N, Paul P (2003). "Reliability properties of reversed residual lifetime". \textit{Communications in Statistics-Theory and Methods}, \textbf{32}, 2031-2042.
\item Navarro J, Franco M, Ruiz J M (1998). "Characterization through moments of the residual life and conditional spacings". \textit{Sankhya}, \textbf{A 60}, 36-48.
\item Pearson K (1895). "Contributions to the Mathematical Theory of Evolution to Skew Variation in Homogeneous Material". \textit{Philos Trans R Soc Lond A}, \textbf{186}, 343-414.
\item Pundir S, Arora S, Jain K (2005). "Bonferroni curve and the related statistical inference". \textit{Statistics \& Probability Letters}, \textbf{75}(2), 140-150.
\item Renyi A (1961). "On Measures of Entropy and Information. In: Proceedings of the 4th Berkeley Symposium on Mathematical". \textit{Statistics and Probability, University of California Press, Berkeley}.
\item Sen S, Maiti S S, Chandra N (2016). "The xgamma distribution: Statistical properties and application". \textit{Journal of Applied Statistical Methods}, \textbf{15}(1), 774-788.
\item Silva F G, Percontini A, Brito E D, Ramos W M, Venancio R, Cordeiro G (2017). "The Odd Lindley-G Family of Distributions". \textit{Austrian Journal of Statistics}, \textbf{46}, 65-87.
\item Tukey J W (1960). "The Practical Relationship between the Common Transformations of Percentages of Counts and Amounts". \textit{Technical Report}, \textbf{36}, Princeton, NJ: Princeton University, Statistical Techniques Research Group.
\item Zakerzadah, H. and Dolati, A., (2010): "Generalized Lindley distribution", Journal of Mathematical Extension, \textbf{3}(2), 13-25.
\item Bakouch, H.S., Al-Zahrani, B.M., Al-Shomrani, A.A., Marchi, V.A., and Louzada, F.
(2012): "An Extended Lindley distribution", Journal of the Korean Statistical Society, \textbf{41}, 75-85.
\item Elbatal, I., Merovci, F., and Elgarhy, M. (2013): "A New Generalized Lindley distribution",
Mathematical Theory and Modeling, \textbf{3}(13), 30-47.
\item Shanker, R., Sharma, S.,and Shanker, R., (2013): "A two-parameter Lindley distribution
for modeling waiting and survival times data", Applied Mathematics, \textbf{4}, 363-368.
\item Ghitany, M., Al-Mutairi, D., Balakrishnan, N. and Al-Enezi, I. (2013): "Power Lindley
distribution and associated inference", Computational Statistics and Data Analysis, \textbf{64}, 20-33.
\item Singh, S.K., Singh, U. and Sharma, V.K. (2014): "The Truncated Lindley Distribution
Inference and Application", Journal of Statistics Applications and Probability, \textbf{3}(2), 219-228.
\item Abouammoh, A.M., Alshangiti, A.M. and , I.E. (2015): "A new generalized Lindley distribution". Journal of Statistical computation and simulation, \textbf{85}(18), 3662-3678.
\item Bouchahed, L. and Zeghdoudi, H. (2018): "A new and unified approach in generalizing the
Lindley's distribution with applications", Statistics in Transition, \textbf{19}(1), 61-74.

\end{enumerate}
\end{document}